\documentclass[twocolumn]{aastex631}

\accepted{June 1, 2025}

\submitjournal{ApJ}

\shorttitle{Refining the HST SBF Calibration}
\shortauthors{Jensen et al.}
\usepackage{amsmath}
\newcommand{\Mbarj}{\ensuremath{\overline{M}_{110}}}
\newcommand{\mbarj}{\ensuremath{\overline{m}_{110}}}
\newcommand{\gminz}{\ensuremath{(g{-}z)}}
\newcommand{\kmsmpc}{km\,s$^{-1}$\,Mpc$^{-1}$}

\begin{document}

\title{The TRGB$-$SBF Project.\ III.\ Refining the HST Surface Brightness Fluctuation \\ Distance Scale Calibration with JWST
}
\author[0000-0001-8762-8906]{Joseph B. Jensen}
\affiliation{Utah Valley University, Orem, Utah 84058, USA}

\author[0000-0002-5213-3548]{John P. Blakeslee}
\affiliation{NSF's NOIRLab, 950 N Cherry Ave, Tucson, AZ 85719, USA}

\author[0000-0003-2072-384X]{Michele Cantiello}
\affiliation{INAF $–$ Astronomical Observatory of Abruzzo, Via Maggini, 64100, Teramo, Italy}

\author[0009-0008-6333-6536]{Mikaela Cowles}
\affiliation{Utah Valley University, Orem, Utah 84058, USA}

\author[0000-0002-5259-2314]{Gagandeep S. Anand}
\affiliation{Space Telescope Science Institute, 3700 San Martin Drive, Baltimore, MD 21218, USA}

\author[0000-0002-9291-1981]{R. Brent Tully}
\affiliation{Institute for Astronomy, University of Hawaii, 2680 Woodlawn Drive, Honolulu, HI 96822, USA}

\author[0000-0002-5514-3354]{Ehsan Kourkchi}
\affiliation{Institute for Astronomy, University of Hawaii, 2680 Woodlawn Drive, Honolulu, HI 96822, USA}
\affiliation{Utah Valley University, 800 W. University Parkway, Orem, UT 84058, USA}

\author[0000-0002-5577-7023]{Gabriella Raimondo}
\affiliation{INAF $–$ Astronomical Observatory of Abruzzo, Via Maggini, 64100, Teramo, Italy}

\begin{abstract}
The TRGB$-$SBF Project team is developing an independent distance ladder using a geometrical calibration of the tip of the red giant branch (TRGB) method in elliptical galaxies that can in turn be used to set the surface brightness fluctuation (SBF) distance scale independent of Cepheid variables and Type~Ia supernovae {(SNe\,Ia)}. 
{The purpose of this project is to measure the local expansion rate of the universe independently of the methods that are most at odds with the theoretically-predicted value of the Hubble-Lema\^itre constant $H_0$, and therefore isolate the influence of potential systematic observational errors.}
In this paper, we use JWST TRGB distances calibrated using the megamaser galaxy NGC~4258 
to determine a new Cepheid-independent SBF zero point for HST.
This new calibration, along with improved optical color measurements from PanSTARRS and DECam, gives an updated value of $H_0 = 73.8\pm 0.7$ (statistical) $\pm 2.3$~(systematic)~\kmsmpc\ that is virtually identical to the 
{SBF Hubble constant measured by \citet{blakeslee2021}.}

\end{abstract}


\section{Introduction \label{sec:intro}}
The current disagreement between the measured expansion rate of the universe and that {inferred} from the {best fit version of the $\Lambda$CDM cosmological model derived from} the cosmic microwave background fluctuations is now greater than 5-$\sigma$ \citep{planck2020, riess2022}.
Resolving this discrepancy (often referred to as the ``Hubble tension'') is {one of the most pressing issues} in cosmology today.
It is not yet clear if the explanation is an accumulation of various systematic errors in local distance measurements or if {modifications to the $\Lambda$CDM model are} required  \citep[e.g.,][]{divalentino2021,CosmoVerse2025}.

 {Most measurements of $H_0$ in the relatively nearby universe favor values around 73 \kmsmpc\ (see the recent compilation by \citealt{CosmoVerse2025}).
The SH0ES team has achieved the highest precision to date using four geometrical methods to anchor the Cepheid+SNe\,Ia distance ladder \citep{riess2022, riess2024a}. 
The Carnegie Chicago Hubble Program (CCHP) Team, in contrast, measured values of $H_0$ that are closer to the CMB+$\Lambda$CDM values \citep{freedman2024, Hoyt2025}. The discrepancy between these two groups, using many of the same targets and similar methods, is most likely due to choices in the sample selection \citep{riess2024b} and not the geometrical anchors themselves. One of the primary purposes of this project is to test that conclusion by providing a measurement of $H_0$ that is completely independent of the Cepheid and SNe\,Ia techniques, and therefore to distinguish between systematic errors in the zero point calibrations and sample selection or analysis differences. If the Hubble tension between measurements from the CMB and $H_0$ measured locally persists with multiple independent techniques, then the possibility of additional physics beyond $\Lambda$CDM must be taken seriously.} 

The TRGB$-$SBF Project seeks to reduce  {or even eliminate some sources of} systematic uncertainty in the first rungs of the distance ladder by developing a new  {and independent} distance scale based on a geometrical calibration of  {the old, metal-rich RGB population in massive elliptical galaxies. The new distance scale is completely independent of the Cepheid and  {SNe\,Ia} distance measurements, and therefore avoids the systematic uncertainties associated with them.} 
To do this, we are  {calibrating SBF distances measured in 14 nearby elliptical galaxies using TRGB distance measurements from the James Webb Space Telescope \citep{2023jwst.prop.3055T}. }
The TRGB$-$SBF Project will eventually link geometrically-calibrated TRGB distances at 15--20 Mpc with SBF distances reaching out to perhaps 300 Mpc or more \emph{using the same telescope (JWST), camera (NIRCam), and filters to minimize systematic uncertainties for each rung of the ladder.}

SBF distances have already been measured in more than 370 galaxies out to about 100~Mpc using the Hubble Space Telescope (HST), including more than 220 in F110W using the WFC3/IR camera.
In this paper, we use the TRGB distances to eight individual galaxies calibrated using the geometrical distance to the megamaser galaxy NGC~4258 \citep{Reid2019,anand2024a} to calibrate the HST SBF distances via TRGB. In the future, the Gaia satellite \citep{Gaia2017,Gaia2018} will add additional geometrical anchors to the TRGB distance scale to further reduce the systematic calibration uncertainty.
This study  {is the third paper in a series to establish an independent distance ladder based on Population II stars. The first two papers determined TRGB distances to multiple galaxies in the Fornax and Virgo clusters \citep{anand2024b, anand2024c}. This paper} presents the first step towards an SBF calibration that is independent of the Cepheid distance scale and connects JWST TRGB to the existing HST SBF distance measurements out to 100~Mpc.

\section{Infrared Surface Brightness Fluctuations with HST \label{sec:sbf}}

The SBF technique is a well-established high-precision $({\lesssim}5$\% per galaxy) distance indicator that results from the spatial fluctuations in luminosity of a galaxy arising from the discrete number of stars per resolution element \citep{tonry1988, blakeslee2009, jensen2015, jensen2021, Cantiello2024}.
While theoretical calibrations of SBF are possible based on stellar population models that predict the absolute luminosities of RGB and AGB stars \citep{Raimondo2005,Raimondo2009}, the precision of such models cannot currently compete with  empirical calibrations based on other reliable distance techniques, including Cepheids and TRGB.

 {Almost all optical and  {infrared (IR)} HST SBF measurements published in the last decade} are based on the calibrations published by \citet{blakeslee2009}, \citet{blakeslee2010}, and \citet{jensen2015,jensen2021}.
 {The foundation of SBF measurements used in these studies are the HST Cepheid distances} to seven galaxies with ground-based SBF distances measured by \citet{tonry2001} and \citet{blakeslee2002}, combined with a large number of overlapping ground-based and HST SBF measurements in the Virgo and Fornax galaxy clusters \citep[see][for direct comparisons]{blakeslee2010}. 
These large surveys provided the statistical foundation for SBF and showed that the intrinsic scatter in the  {absolute fluctuation magnitudes (and therefore the SBF distance moduli)} is ${\sim\,}0.06$~mag (optical) to ${\sim\,}0.1$~mag (near-IR), after correction for stellar population variations using optical colors \citep{blakeslee2009, jensen2015}.  {The current project seeks to replace the Cepheid calibration with TRGB for the purpose of isolating systematic uncertainties in the various distance measurement techniques.}

Fluctuations are  {more prominent} at near-IR wavelengths than in the optical, and are measurable to ${\gtrsim}80$~Mpc in a single HST orbit.
 {The SBF method} is much more efficient than Cepheids or  {SNe\,Ia} for  {measuring} distances out to 100~Mpc  {because only one imaging observation is required, and SBF} achieves comparable precision  {in much less observing time}.
The majority of the SBF distances beyond 40~Mpc useful for measuring $H_0$ were made using NICMOS  \citep{jensen2001} and WFC3/IR \citep{cantiello2018,blakeslee2021}. 
The most recent WFC3/IR F110W SBF calibration was determined using 11 galaxies in the Fornax and Virgo galaxy clusters, and was corrected for stellar population variations using ACS $(g_{475}{-}z_{850})$ colors \citep{jensen2015}.
Updates to the \citet{jensen2015} calibration to include better Galactic foreground extinction corrections and an improved distance to the LMC  were made as described by \citet{cantiello2018} and \citet{jensen2021}.

The majority of the HST IR SBF observations published between 2015 and 2021 were part of HST programs GO-14219 (PI J.~Blakeslee), GO-14654 (PI P.~Milne), and GO-15625 (PI J.~Blakeslee). \citet{jensen2021} published SBF distances to 63 galaxies from these programs, including 25  {SNe\,Ia} host galaxies. 
Three more HST programs  {have recently been completed}. GO-16262 (PI R.~Tully), GO-17436 (PI J.~Jensen), and GO-17446 (PI P.~Milne) have already observed ${\gtrsim}164$ more galaxies out to ${\gtrsim}85$~Mpc, a sample that more than doubles the number of  {SNe\,Ia} host galaxies with high-quality SBF distances (now 51). To make the most of these new datasets, in this paper we update the HST SBF calibration using the new JWST TRGB distances.

The SBF distances published by \citet{jensen2021} were used by \citet{blakeslee2021} to measure a value of the Hubble 
constant of $H_0 = 73.3 \pm 0.7$ (statistical) $ \pm 2.4$ (systematic) ~km\,s$^{-1}$\,Mpc$^{-1}$. 
The subsample of  {25 early-type SNe\,Ia host galaxies with HST SBF distances provided an alternative calibration of SNe\,Ia in the Hubble flow:} \citet{garnavich2023} measured a value of 
$H_0 = 74.6 \pm 0.9$  {(statistical)} $\pm 2.7$  {(systematic)}~km\,s$^{-1}$\,Mpc$^{-1}$ using SBF distances  {as an intermediate step between Cepheids and SNe\,Ia.
\citet{garnavich2023} used the Pantheon+ \citep{pantheonplus} light curves to make the SNe\,Ia distance measurements.}
 {When the light curve fitting parameters 
were corrected to better match the fast-declining SNe\,Ia preferentially} found in the early-type galaxies, the value of $H_0$ became $73.3\pm1.0\pm2.7$ km\,s$^{-1}$\,Mpc$^{-1}$.

 {The uncertainties in most of the previous SBF $H_0$ measurements
are dominated by the random uncertainties associated with the intrinsic scatter among the limited number of calibrator galaxies (seven), which becomes a systematic uncertainty when applied to the next rung on the distance ladder (SBF in this case). The calibrator galaxies are those in which both Cepheids and SBF have been measured; Cepheids are most commonly detected in spiral galaxies, while SBF can only be measured in early-type S0 and elliptical galaxies.}
 {The systematic SBF distance uncertainty from the Cepheid calibration} reported by \citet{blakeslee2021} included 0.028~mag from the Cepheid zero point (calibrated using the LMC), 0.08~mag from linking Cepheids in spiral galaxies to SBF measurements in the bulges of those galaxies, and 0.03~mag from connecting the optical ACS and WFC3/IR SBF, for a total of 0.09~mag (4.2\% in distance  {or 3.1~\kmsmpc\ in $H_0$}). 
 {Even when combined with HST TRGB measurements, the systematic uncertainty of 2.7~\kmsmpc\ } was $3{\times}$ the statistical uncertainty for the sample of 63 galaxies \citep{blakeslee2021}. 

\section{Re-calibrating the HST WFC3/IR SBF Distance Scale using TRGB \label{directcal}}

JWST observations of the TRGB in 14 nearby elliptical galaxies (GO-3055; \citealt{2023jwst.prop.3055T}) provide a new and direct calibration of IR SBF based on the geometrical megamaser distance to NGC~4258 \citep{anand2024a, Reid2019}. 
We will eventually use JWST to determine $H_0$ directly by measuring SBF distances out to ${\sim}\,300$~Mpc, but building a database of distant SBF measurements will take time, and may never reach the quantities already observed with HST over the last 30 years (more than $370$ galaxies, including SBF distances from NICMOS, ACS, and WFC3/IR). 
To leverage the full potential of the HST database, we need to connect the new JWST TRGB distances to the existing HST SBF dataset.
Three of the JWST TRGB target galaxies in the Fornax cluster and five in  {the Virgo cluster} have also been observed with WFC3/IR, as shown in Table~\ref{tab:calibrationdata}.
The new JWST TRGB measurements of these galaxies \citep{anand2024b,anand2024c} provide us with an opportunity to reduce systematic uncertainties in SBF distances originating  {from} the Cepheid calibration. Using a direct geometrical calibration of TRGB from NGC~4258 \citep{Reid2019, anand2024a} and SBF measurements in the \emph{same} galaxies, using the \emph{same} telescope, camera, and filters, we can eliminate some sources of uncertainty (e.g., Cepheid metallicity corrections in different galaxy types and positional offsets between spirals and ellipticals in the calibration clusters).

\subsection{Updated Calibration}

 {In this paper we re-calibrated the HST IR SBF distance scale by replacing the Cepheid-based optical SBF distances} from \citet{jensen2015} with the new JWST-based TRGB distances to the eight galaxies in common between the HST and JWST SBF calibration data sets.
The SBF calibration of the absolute fluctuation magnitude  {in the F110W filter, denoted with the symbol \Mbarj,} depends on the distance scale zero point and a slope with galaxy color, which corrects \Mbarj\ for variations in stellar population age and metallicity \citep[e.g.,][]{jensen2003}.
\citet{jensen2015} found that the  {rms scatter among the calibration galaxies was 25\% smaller} when using the average distance to the cluster instead of individual optical SBF distances.\footnote {Whether or not the individual optical SBF distances will give a better calibration depends on the magnitude of the uncertainties relative to the cluster depth. When individual observational uncertainties are larger than the cluster depth, averaging first produces a lower rms in the calibration scatter; when they are smaller than the cluster depth, then using the individual distances will result in a lower scatter. In the case of the ACS SBF measurements in Virgo and Fornax, the individual uncertainties and the cluster depth were comparable (see \citealt{jensen2015} for details).} 
We can do the same here, and by using the mean cluster distances from the JWST TRGB measurements we include four more galaxies in  {the Fornax cluster} and two more in  {the Virgo cluster}, for a total of 14 galaxies in our new calibration (see Table~\ref{tab:calibrationdata}).

\begin{deluxetable*}{lccccccc}[ht]
\tabletypesize{\footnotesize}
\tablewidth{0pt}
\caption{SBF Calibration Data\label{tab:calibrationdata}}
\tablehead{
\colhead{Galaxy} & \colhead{$(m{-}M)_{\rm TRGB}$\tablenotemark{}} & \colhead{TRGB $\sigma_{\rm tot}$\tablenotemark{}} & \colhead{$(m{-}M)_{\rm clust}$\tablenotemark{}} & \colhead{\mbarj$_{,0}$\tablenotemark{}} & \colhead{DECam $(g{-}z)_0$\tablenotemark{}} & \colhead{PS $(g{-}z)_0$\tablenotemark{}} & \colhead{HST}  \\
& \colhead{(mag)} &\colhead{(mag)} &\colhead{(mag)} &\colhead{(AB mag)}&\colhead{(AB mag)}&\colhead{(AB mag)}&\colhead{Program}
\\
 &(1) & (2) & (3) & (4) & (4, 5) & (4) & (6) 
}

\startdata
IC~2006  & \nodata        & \nodata & $31.424\pm0.057$ & $28.65 \pm 0.021$ & $1.274\pm0.014$ & \nodata &11712 \\
NGC~1344 & \nodata        & \nodata & $31.424\pm0.057$ & $28.46 \pm 0.019$ & $1.201\pm0.012$ & \nodata & 11712 \\
NGC~1374 & \nodata        & \nodata & $31.424\pm0.057$ & $28.58 \pm 0.025$ & $1.226\pm0.012$ & \nodata & 11712 \\
NGC~1375 & \nodata        & \nodata & $31.424\pm0.057$ & $28.26 \pm 0.021$ & $1.075\pm0.012$ & \nodata & 11712\\
NGC~1380 & $31.397\pm0.034$ & 0.072 & $31.424\pm0.057$ & $28.60 \pm 0.022$ & $1.283\pm0.012$ & \nodata & 11712\\
NGC~1399 & $31.511\pm0.036$ & 0.073 &$31.424\pm0.057$  & $28.84 \pm 0.023$ & $1.363\pm0.012$ & \nodata & 11712 \\
NGC~1404 & $31.364\pm0.034$ & 0.072 & $31.424\pm0.057$ & $28.73 \pm 0.024$ & $1.325\pm0.012$ & \nodata & 11712 \\
NGC~4458 & \nodata        & \nodata & $31.055\pm0.086$ & $28.05 \pm 0.023$ & $1.149\pm0.012$ & $1.170\pm0.023$ & 11712 \\
NGC~4472 (M49)& $31.091\pm0.032$ & 0.071 & $31.055\pm0.086$ & $28.43 \pm 0.021$ & $1.383\pm0.012$ & $1.380\pm0.011$ & 11712\\
NGC~4489 & \nodata        & \nodata & $31.055\pm0.086$ & $27.87 \pm 0.019$ & $1.126\pm0.013$ & $1.109\pm0.024$ & 11712 \\
NGC~4552 (M89)& $30.933\pm0.041$ & 0.075 & $31.055\pm0.086$ & $28.31 \pm 0.022$ & $1.349\pm0.013$ & $1.366\pm0.012$ & 15082 \\
NGC~4636 & $31.120\pm0.035$ & 0.072 & $31.055\pm0.086$ & $28.36 \pm 0.021$ & $1.332\pm0.013$ & $1.330\pm0.013$ & 17446 \\
NGC~4649 (M60)& $31.061\pm0.034$ & 0.072 & $31.055\pm0.086$ & $28.52 \pm 0.021$ & $1.426\pm0.013$ & $1.423\pm0.011$ & 11712 \\
NGC~4697 & $30.330\pm0.036$ & 0.073 & $30.330\pm0.073$ & $27.52 \pm 0.021$ & $1.271\pm0.013$ & $1.271\pm0.011$ & 15226 \\
\enddata 
\tablenotetext{}{(1) JWST TRGB distances from \citet{anand2024b} and \citet{anand2024c}. JWST data are available through MAST, \dataset[DOI: 10.17909/z9ch-wk24]{https://doi.org/10.17909/z9ch-wk24}.}
\tablenotetext{}{(2) Total uncertainty for the TRGB distance including systematic uncertainty.}
\tablenotetext{}{(3) Average cluster TRGB distance moduli for Virgo (including NGC~4636) and Fornax, except for NGC~4697, which is in the foreground \citep{anand2024b,anand2024c}. The uncertainties are the cluster depths from the ACS Virgo and Fornax Cluster surveys \citep{blakeslee2009}, 0.085 and 0.053~mag respectively, added in quadrature with the uncertainty in the cluster averages of the TRGB distances (0.02~mag for the three TRGB galaxies in Fornax and 0.013~mag for the seven Virgo).}
\tablenotetext{}{(4) SBF magnitudes and optical colors have been corrected for Galactic extinction using \citet{SF2011} and the NED extinction calculator \dataset[DOI: 10.26132/NED5] {https://doi.org/10.26132/NED5}. {Extinction corrected values are indicated with a subscript 0.} All fluctuation magnitudes and colors were measured in two annular region around each galaxy from 8.2 to 16.4 arcsec and 16.4 to 32.8 arcsec in radius. The values shown are the weighted averages of the two measurements.}
\tablenotetext{}{(5) DECam $(g{-}z)_0$ colors have been transformed to the PanSTARRS photometric system using Equation~1 \citep{decamphotometry2019} to simplify comparison.}
\tablenotetext{}{(6) WFC3/IR data for these galaxies are available through MAST, \dataset[DOI: 10.17909/303n-6g35]{https://doi.org/10.17909/303n-6g35}.}

\end{deluxetable*}

The direct overlap between JWST {(TRGB)} and HST {(Cepheid)} calibration {samples} includes three galaxies in the Fornax cluster and five in  {the Virgo cluster} (including NGC~4636 and NGC~4697 in peripheral in-falling regions).  
Using the combined average  {TRGB distance modulus to the Virgo cluster of 
$31.043\pm0.034$~(statsitical)~$\pm0.063$~(systematic) mag
\citep{anand2024b,anand2024c} and the {Cepheid-based} ACSVCS survey average distance modulus of $31.092 \pm 0.013$~mag \citep{mei2007, blakeslee2009}, which was used to calibrate the previous IR SBF measurements {\citep{jensen2021}}, the weighted mean difference is $\Delta\mu_{\rm TRGB-Ceph} = -0.067\pm 0.031$~mag \citep{anand2024c}. 
For Fornax, the difference between the mean ACS SBF distance modulus of 31.51~mag \citep{blakeslee2009} used for the \citet{jensen2015} IR SBF calibration and the mean distance from three JWST TRGB measurements gives a difference of $\Delta \mu = -0.086\pm0.077$~mag, where the  {uncertainty is the result of} the scatter among the three galaxies.}

In both  {the Virgo and Fornax clusters}, the TRGB-based distances appear to be 3 to 4\% shorter, with $\sim\,$1 to 2$\sigma$ significance, than the Cepheid-based SBF distances used for calibrating the IR SBF distance scale \citep{jensen2015}. 
 {While we do not yet know the reason for the marginally-significant difference between the TRGB and Cepheid distances to the Virgo and Fornax clusters,} the implications for the Hubble constant measured using the new TRGB zero point calibration alone is that $H_0$ would be 
larger than the values published by \citet{blakeslee2021} and \citet{garnavich2023},  {which both use} the \citet{jensen2021} SBF calibration. 

The SBF measurements were updated since the original calibration published by \citet{jensen2015}, and for this study we have included three additional calibration galaxies (NGC~4552, NGC~4636, and NGC~4649).
The new  {apparent SBF magnitude} \mbarj\ measurements for the calibrators are consistent with the procedures we used to measure \mbarj\ in the sample from which we derived $H_0$ \citep{jensen2021,blakeslee2021}. 
The average difference between the  SBF magnitudes \mbarj\ from the \citet{jensen2015} calibration paper and the present values for the five calibrator galaxies in common is 0.072 mag with a standard deviation of 0.03~mag. 
 {\citet{jensen2021} introduced} a number of improvements to the SBF analysis  {(e.g., improvements to the PSF normalization, modernization of the LMC zero point calibration for Cepheids, and improvements to the procedure to remove SBF signal resulting from undetected globular clusters)} that were previously applied to the sample used to determine $H_0$ \citep{blakeslee2021}, and those have been applied to these new measurements. We previously identified an offset of $0.05\pm0.02$~mag to the 2015 calibration values due to the PSF normalization \citep{cantiello2018}. Given the small sample size in this study, the 0.072~mag offset is considered to be statistically consistent with the previously measured 0.05~mag offset. Further assessment of the reliability of the \mbarj\ measurements will require additional overlap between calibration samples.

Optical galaxy colors are critical for the SBF method as they are necessary to correct SBF amplitudes for intrinsic variations in the ages and metallicities of the stellar populations in elliptical galaxies \citep{jensen2003, blakeslee2009, jensen2015}. 
The original IR SBF calibration was based on the ACS $(g_{475}{-}z_{850})$ colors, which are not generally available for the vast majority of the WFC3/IR observations in the HST archive.
\citet{jensen2021} transformed PanSTARRS (PS) $(g{-}z)$ colors \citep{chambers2019panstarrs1surveys} to the ACS photometric frame for computing distances.
For this study, we remeasured all the PanSTARRS $(g{-}z)$ colors and determined the SBF calibration directly in that system.
We found that for galaxies with significant color gradients, the PanSTARRS \gminz\ colors used by \citet{jensen2021} for computing SBF distances---measured in the high-$S/N$ ratio central regions---did not always match the colors in the regions where the SBF was measured.
For typical massive galaxies with slightly redder cores and bluer halos, correcting the colors to match the SBF regions leads to larger distances, by as much as 5\% in some cases, and hence this revision tends to reduce the value of $H_0$. 
For the 61 galaxies in the \citet{jensen2021} dataset with PanSTARRS photometry, which are used again in this paper to recompute $H_0$, the color shift was measured to be $-0.023\pm0.019$~mag (bluer). The average photometric uncertainty in \gminz\ is 0.02~mag for PanSTARRS, which includes the uncertainty in the extinction correction \citep{SF2011} added in quadrature.
Given that the historical slope of the relationship between SBF \Mbarj\ and \gminz\ is ${\sim\,}2$  \citep{jensen2015}, the expected shift in distance modulus is approximately 0.05~mag or 2.3\% in distance, counteracting in part the increase in $H_0$ expected from the new TRGB zero point.

The original SBF calibration of \citet{jensen2015} included equal numbers of Fornax and Virgo cluster galaxies, which were calibrated using ACS optical photometry. The JWST TRGB sample includes three galaxies in the Fornax cluster. The PanSTARRS-1 survey does not reach far enough south to include Fornax, so the current TRGB calibration is limited to five galaxies in  {the Virgo cluster}.
 {Surveys using DECam on the Blanco Telescope at CTIO provide a high-$S/N$ alternative to PanSTARRS for both the Fornax and Virgo clusters. A number of imaging surveys (e.g., DESI and DECaLS) using DECam are now publicly available as part of the Legacy Surveys \citep{decamphotometry2019}.}\footnote{\url{https://datalab.noirlab.edu/sia.php}}
We retrieved data in the Legacy Survey DR10  release\footnote{\url{https://www.legacysurvey.org/dr10/description/}} through the server at the National Energy Research Scientific Computing Center (NESRC).
We combined the archival images (``bricks'') for 23 target galaxies in the \citet{jensen2021} sample and measured their \gminz\ colors. Uncertainties were determined by similarly combining variance frames.
The DECam Legacy Survey photometric uncertainties are 0.0073~mag for $g$ and $z$, which were combined in quadrature with each other and with the uncertainty in the foreground extinction correction from \citet{SF2011}. The mean uncertainty in \gminz\ is 0.011~mag.

\begin{figure}
\begin{center}
\includegraphics[scale=0.85]{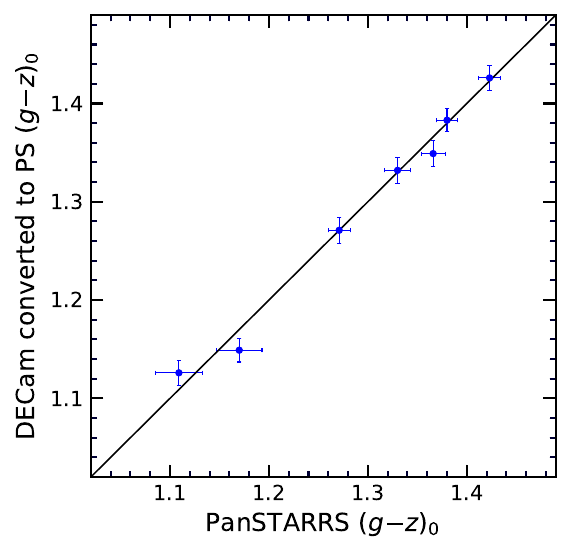}
\caption{Comparison of the  {extinction-corrected} PanSTARRS \gminz$_0$\ colors with those transformed from the DECam colors to PanSTARRS using \citet{decamphotometry2019} for the seven calibrator galaxies with color data in both surveys. The black line denotes the one-to-one relation and is not a fit to the data.
\label{fig:transformation}}
\end{center}
\end{figure}

\citet{decamphotometry2019} provide a transformation between PanSTARRS and DECam \gminz\ values as follows:
\begin{equation}
\begin{split}
(g{-}z)_{\rm PS} = (g{-}z)_{\rm DECam} + 0.02521 - 0.11294(g{-}i) + \\ + 0.01796(g{-}i)^2 -0.00285(g{-}i)^3 
\end{split}
\end{equation}

\noindent where $(g{-}i)$ is the same in both photometric systems.
For the seven galaxies  with \gminz\ colors in both systems, the predicted vs.\ computed values of \gminz\ differ by $0.002\pm 0.013$~mag (Fig.~\ref{fig:transformation}).

Since DECam \gminz\ colors were measured for both the Fornax and Virgo cluster samples, we used the DECam colors for the calibration, but transposed to the PanSTARRS photometric system because PanSTARRS was used for the published galaxy sample used to measure $H_0$ \citep{jensen2021,blakeslee2021}. The values listed in Table~\ref{tab:calibrationdata} are in the PanSTARRS photometric system and have an additional uncertainty of 0.013~mag added in quadrature to account for the observed scatter in the transformation. 

\begin{figure}
\begin{center}
\includegraphics[scale=0.85]{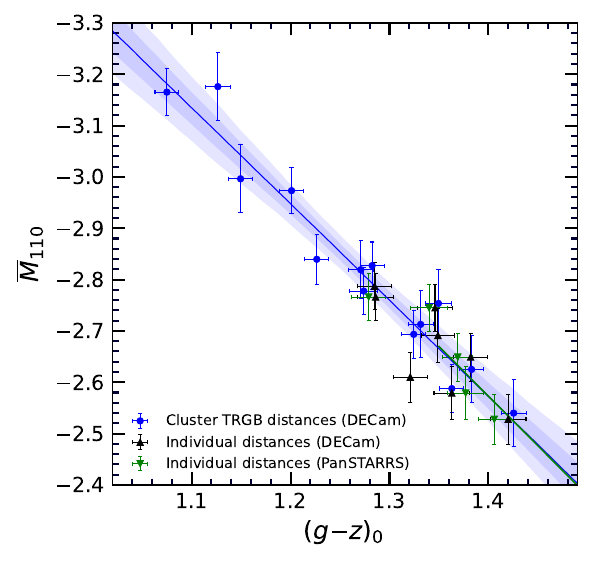}
\caption{Absolute fluctuation magnitude \Mbarj\ as a function of  {the extinction-corrected DECam galaxy color \gminz$_0$} converted to the PanSTARRS photometric system for the eight galaxies with TRGB distances \citep{anand2024b,anand2024c} plus six additional galaxies in the Fornax and Virgo clusters previously used to calibrate F110W SBF (blue points). 
The fit shown was performed using the ``emcee'' package in Python 
 {\citep{emcee2013} 
that we used to} iteratively account for uncertainties in both axes.
The blue line is the best fit to the blue points;
the blue envelopes mark the 1-$\sigma$ and 2-$\sigma$ uncertainties in slope and intercept. 
The absolute magnitudes calculated using the individual TRGB distance moduli are also plotted for five of the galaxies with PanSTARRS colors (green points) and eight galaxies with DECam colors (black points) to show that slopes derived from the individual and cluster distances are consistent. 
\label{fig:Mbar}}
\end{center}
\end{figure}

The eight galaxies for which JWST TRGB distances were individually measured (black points in the right half of Figure~\ref{fig:Mbar}) provide a zero point  {with an accuracy of} 0.018~mag but a relatively weak constraint on the slope due to the limited range in color covered by the JWST sample (the slope is $1.7\pm 0.3$ in the PanSTARRS \gminz\ photometric system).
To determine a more robust measurement of the slope, we adopted the mean cluster TRGB distance for the Virgo and Fornax clusters, and used those distances to compute \Mbarj\ for a larger sample of HST Virgo and Fornax calibrators from \citet{jensen2015} that include a wider range of galaxy color.
When we include six additional galaxies and calculate \Mbarj\ using mean TRGB cluster distances (see Table~\ref{tab:calibrationdata}) we get a tighter value for the slope of $1.86\pm0.16$ with a reduced $\chi_\nu^2=0.74$ per degree of freedom. 
The fact that the reduced $\chi_\nu^2$ is less than one suggests that the uncertainties in the absolute magnitude may be slightly overestimated, but
there is still a 30\% probability of measuring this value of $\chi_\nu^2$ for the number of degrees of freedom in our sample, indicating general consistency.
Regardless of which TRGB distances (and slope) are used, the zero point at a color of $(g{-}z) = 1.30$ is consistently $-2.76$ with an uncertainty below 0.02 mag.

The calibration calculated using the mean TRGB cluster distances (Fig.~\ref{fig:Mbar}) is:
\begin{equation}
   \Mbarj = (1.86\pm0.16)[(g{-}z)_{\rm PS} {-} 1.30] + (-2.760{\pm}0.016)
\end{equation}
\noindent where the color is measured in the PanSTARRS \gminz\ photometric system. 
The calibration in the original DECam \gminz\ system is:
\begin{equation}
   \Mbarj = (1.75\pm0.14)[(g{-}z)_{\rm DC} {-} 1.40] + (-2.728{\pm}0.017).
\end{equation}

\subsection {Updated SBF Distances and $H_0$}

To assess the impact of the new calibration on the IR SBF distance scale, we applied the new TRGB calibration and PanSTARRS colors to the published fluctuation magnitudes for 61 galaxies from \citet{jensen2021} and \citet{blakeslee2021}, and recomputed their distances. 
We also measured DECam \gminz\ colors for 23 galaxies in the distant sample (the rest were too far north to be included in the DECam Legacy Surveys).
The new TRGB zero point and updated colors have opposing effects on the measured distances, and the new calibration implies almost no change in the SBF distances from \citet{jensen2021} or the resulting values of $H_0$ from \citet{blakeslee2021} or \citet{garnavich2023}.

Overall, the new calibration results in an average offset in the distance modulus of $0.011\pm0.005$~mag with RMS scatter of 0.04~mag. This implies a change in the distance scale of a factor of $0.995\pm0.002$ relative to the 2021 distances, meaning that our new distances are on average $0.5\%\pm0.2\%$ closer, and, consequently, the value of $H_0$ would be 0.5\% larger.
We obtain very similar results if we choose to use only the eight calibrators for which direct TRGB distances are available (distance scale factor of $1.000\pm0.003$) or the five Virgo calibrators with PanSTARRS colors (scale factor $1.007\pm0.002$), 
even though the slope of the \Mbarj--color relation is not as well constrained. 
The result is robust to the selection of galaxy colors (DECam or PanSTARRS) and to the choice of individual or cluster TRGB distances.
For example, if we use only DECam colors with no translation to PanSTARRS, and only use the subset of 23 galaxies in the $H_0$ sample with DECam colors, we get a mean distance modulus offset of $0.002\pm0.013$~mag, or a distance scale factor of $0.999\pm0.006$. 

\begin{figure}
\begin{center}
\includegraphics[scale=0.4]{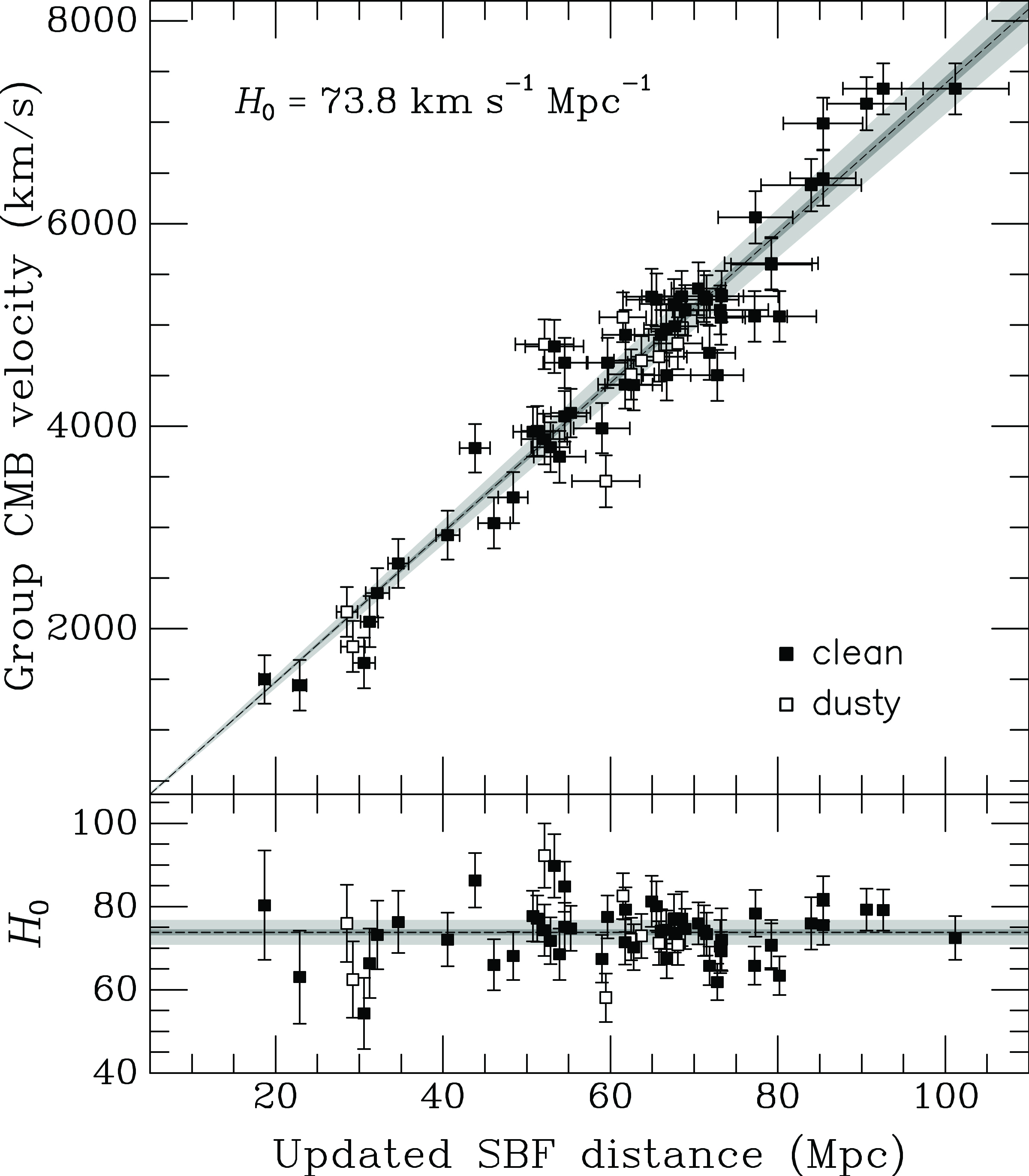}
\caption {Hubble diagram using distances derived from the updated SBF zero point calibration from the TRGB measurements and improved optical color measurements. Velocities and SBF apparent magnitudes are from \citet{jensen2021}. The solid symbols are those for which there is no evidence of dust or other sprial structure in the SBF analysis region. 
This plot is directly comparable to the left panel of Fig.~1 from \citet{blakeslee2021}.
\label{fig:hubbleplot}}
\end{center}
\end{figure}

 {To determine the updated value of $H_0$, we combined the revised SBF distances described above with the CMB-frame group/cluster velocities published by \citet{jensen2021} without refitting for $H_0$. Specifically, we took the mean change in distance computed from 61 galaxies, using the new TRGB zero point and including the updated \gminz\ colors, and applied it to the value of $H_0$ from \citet{blakeslee2021}. } 
The resulting value of $H_0$ using the updated zero point and the published calibration slope gives $73.81\pm0.7$~\kmsmpc\ instead of $73.44\pm0.7$~\kmsmpc. 
Since we are only changing the calibration and \emph{not} the input \mbarj\ fluctuation magnitudes or velocities, the statistical error in the value of $H_0$ is unchanged.  {The updated Hubble diagram is shown in Fig.~\ref{fig:hubbleplot}. The Hubble velocities used to compute this value of $H_0$ and for Figure~\ref{fig:hubbleplot} are in the CMB frame and assigned to groups or clusters by \citet{Tully2015}.}

 {A similar 0.5\% increase would result for the other values of $H_0$ calculated from IR SBF distances or velocities \citep{blakeslee2021, garnavich2023}. 
For example, \citet{blakeslee2021} calculated a value of $H_0 = 73.8$~\kmsmpc\ using the 2M++ redshifts and peculiar velocity field derived from the density model calculated by \citet{2M++2015}; the 2M++ value of $H_0$ using the new zero point is $74.2$~\kmsmpc.
}

The reassessment of the calibration was completed independently without reference to the resulting distances to avoid any confirmation bias in the value of $H_0$.

\subsection{Systematic Uncertainties}

The SBF distance scale is primarily limited by systematic uncertainties.
The Cepheid calibration of SBF injected a 0.09~mag systematic uncertainty \citep{blakeslee2021} that we remove here and replace with that of the NGC~4258 megamaser and JWST TRGB calibration.
The systematic uncertainty associated with the maser distance to NGC~4258 is 1.48\% (0.032~mag, \citealt{Reid2019}). 
The TRGB measurement uncertainty of 0.033~mag for NGC~4258 from \citet{anand2024a} acts as an additional systematic error in the TRGB distances of other galaxies that are tied to it. This source of systematic error cannot be reduced further without  observations of additional fields around NGC~4258 or other geometrical anchors besides NGC~4258.
The link from TRGB to SBF contains the zero point uncertainty of 0.016~mag from Equation~2 above.
Combined in quadrature, these add up to 0.049~mag. 
The intrinsic scatter in the properties of the TRGB feature (and how it is measured) has not been fully quantified. \citet{anand2024a} estimated a scatter of 0.02~mag in their study, and estimates from \citet{riess2024b} for a sample of spiral galaxies observed in multiple filters with JWST suggest larger values from 0.01 to 0.08~mag. 
The elliptical galaxies we observed span a smaller range in age and metallicity and should have modest intrinsic scatter; we conservatively include another 0.04~mag in our error budget.

Taken together, these terms give a total systematic uncertainty on the TRGB calibration of HST SBF of 0.063~mag, or 2.9\% in distance. Combining this with the estimated 1\% uncertainty in the overall velocity flow of the volume covered by our sample \citep[see][]{blakeslee2021}, the total systematic uncertainty in $H_0$ becomes 3.1\%, or 2.3 km\,s$^{-1}$\,Mpc$^{-1}$ in $H_0$.  {While this is only slightly better than the systematic uncertainty reported by \citet{blakeslee2021}, it provides strong confirmation that systematic errors in the SBF distance scale zero point are accurately measured, given that the sources of  systematic uncertainty in the two measurements are very different (i.e., Cepheids vs.\ TRGB). The new TRGB-SBF distance calibration} will improve further with additional geometrical anchors tied to TRGB with Gaia and with more links between JWST TRGB and HST SBF.

 {
\section{Distance to the Coma Cluster}
The Coma Cluster is an important benchmark for studies of the distance scale. \citet{Said2024} used the \citet{jensen2021} distance to a single galaxy in this cluster (NGC~4874, $d = 99.1 \pm 5.8$ Mpc) to calibrate Fundamental Plane (FP) distances derived using spectroscopic and imaging data for thousands of galaxies from the DESI survey. With this calibration, they determined a value of $H_0 = 76.05 \pm 0.35$ (statistical) $\pm 0.49$ (systematic from FP)$ \pm 4.86$ (statistical from the calibration) \kmsmpc. 
Since only one galaxy was used, the average distance offset from the previous section is not appropriate for estimating the impact of our new calibration on the FP value of $H_0$. The newly measured distance to NGC~4874, using the updated TRGB zero point and revised optical color estimate, is $d = 101.2\pm6.4$~Mpc, which is 2\% larger than the 2021 measurement, corresponding to a slightly lower value of $H_0 = 74.5$~\kmsmpc\ for the DESI FP distances.
\citet{DESI-Scolnic2025} compared the DESI FP results to several other distance measurements to the Coma cluster, including SNe Ia, to conclude that the DESI results are inconsistent with the Planck CMB + $\Lambda$CDM model at 3$\sigma$. The revised SBF distance to Coma is still far too small to be consistent with the valued of $H_0$ predicted by the early-universe measurements.
In the near future we will have a sample of over 50 galaxies with JWST-based SBF distances in the Coma cluster \citep{2024jwst.prop.5989J} to better constrain its distance.
}

\section{Conclusions}
The re-calibration of the WFC3/IR F110W SBF distances \citep{jensen2021,blakeslee2021} using the new JWST-based TRGB zero point results in a value of $H_0 = 73.8\pm 0.7 \pm 2.3$~\kmsmpc\ that is entirely independent of Cepheid and  {SNe\,Ia} distances. Our measurement
agrees closely with the previous SBF-based values of $H_0$ calibrated using Cepheids from \citet{blakeslee2021} and \citet{garnavich2023}; it is also consistent with the SH0ES distance ladder based on Cepheids and  {SNe\,Ia} \citep{riess2022}.
It is 2.6$\sigma$ discrepant with the value of $H_0$ from Planck for the $\Lambda$CDM cosmology \citep{planck2020}  {and therefore provides independent evidence for the Hubble tension.}
The purpose of this study has been to explore the impact of the new JWST TRGB distances on the IR SBF distance scale using published HST SBF data; a future study will include the many new HST SBF measurements  acquired since \citet{jensen2021} and \citet{blakeslee2021} were published.

This updated calibration is a first step in establishing a new distance ladder with JWST that will be independent of the traditional rungs that use the LMC distance, Cepheids and  {SNe\,Ia}.
We are optimistic that additional galaxies will be observed by both JWST and HST, increasing the overlap between the two telescopes both for TRGB and SBF measurements, and further reducing systematic uncertainties.

JWST imaging observations of 39 targets in the Coma cluster (likely including ${\sim\,}50$ or more elliptical galaxies) have been scheduled for Cycle 3 (GO-5989; \citealt{2024jwst.prop.5989J}) and will provide a robust determination of the calibration slope in a cluster that is foundational to the distance scale.
Additional geometrical anchors securing the absolute magnitude of the TRGB will further enhance the precision of the SBF method. The combination of HST and JWST SBF measurements will create a new distance ladder based solely on old, metal-rich populations in early-type galaxies and free from many of the uncertainties in the Cepheid+ {SNe\,Ia} distance ladder  {the most important ones of which are crowding \citep[e.g.,][]{riess2023}, Cepheid metallicity effects \citep[e.g.,][]{Madore2025, Bhardwaj2024}, and the effects of dust extinction on both SNe\,Ia and Cepheids \citep[e.g.,][]{Brout2023}.}

 {Reducing the systematic uncertainty in the TRGB-SBF distance scale will require additional geometrical anchors besides NGC~4258, particularly using Gaia \citep{anand2024c, Gaia2016} to precisely measure RGB, Horizontal Branch, and RR Lyrae stellar distances where the TRGB can also be measured. 
Improvements from an expanded SBF sample are also forthcoming. Three HST programs have recently been completed and add more than 150 new IR SBF measurements.
JWST observations of the Coma cluster \citep{2024jwst.prop.5989J} will soon establish a foundation for the SBF calibration that will extend to much larger distances than is possible with HST, and 11 additional TRGB targets have been approved for the next cycle \citep{2025jwst.prop.7034T}. The updated SBF calibration will be applied to JWST observations of galaxies reaching 250~Mpc in cycle 4 (GO-7113).
These projects will further reduce the random and systematic uncertainties on $H_0$ measured using the extensive HST WFC3/IR SBF dataset and new observations with JWST. 
}

\medskip
\begin{footnotesize}

\emph{Acknowledgments.} This research is based on observations made with the NASA/ESA Hubble Space Telescope obtained from the Space Telescope Science Institute, which is operated by the Association of Universities for Research in Astronomy, Inc., under NASA contract NAS 5–26555. The results presented here used data associated with HST programs GO-11711, 11712, 12450, 14219, 14654, 15265, 14771, 14804, and 15329. J.~Jensen and M.~Cowles acknowledge partial support from STScI under grants HST-GO-16262 and JWST-GO-03055. This work also used observations made with the NASA/ESA/CSA James Webb Space Telescope 
program GO-3055. 
J.~Blakeslee is supported by NOIRLab, which is managed by the Association of Universities for Research in Astronomy (AURA) under a cooperative agreement with the U.S. National Science Foundation. M.~Cowles acknowledges support from The Barry Goldwater Scholarship and Excellence in Education Foundation and the National Science Foundation. M. Cantiello acknowledges support from the ASI-INAF agreement ``Scientific Activity for the Euclid Mission'' (n.2024-10-HH.0; WP8420) and from the INAF ``Astrofisica Fondamentale'' GO/GTO-grant 2024 (P.I. M. Cantiello). 
G. S. Anand acknowledges financial support from JWST GO–3055.

This project used data obtained with the Dark Energy Camera (DECam), which was constructed by the Dark Energy Survey (DES) collaboration. Funding for the DES Projects has been provided by the DOE and NSF (USA), MISE (Spain), STFC (UK), HEFCE (UK), NCSA (UIUC), KICP (U. Chicago), CCAPP (Ohio State), MIFPA (Texas A\&M), CNPQ, FAPERJ, FINEP (Brazil), MINECO (Spain), DFG (Germany) and the Collaborating Institutions in the Dark Energy Survey, which are Argonne Lab, UC Santa Cruz, University of Cambridge, CIEMAT-Madrid, University of Chicago, University College London, DES-Brazil Consortium, University of Edinburgh, ETH Zürich, Fermilab, University of Illinois, ICE (IEEC-CSIC), IFAE Barcelona, Lawrence Berkeley Lab, LMU M\"unchen and the associated Excellence Cluster Universe, University of Michigan, NSF NOIRLab, University of Nottingham, Ohio State University, OzDES Membership Consortium, University of Pennsylvania, University of Portsmouth, SLAC National Lab, Stanford University, University of Sussex, and Texas A\&M University.

This research uses services or data provided by the Astro Data Lab, which is part of the Community Science and Data Center (CSDC) Program of NSF NOIRLab. NOIRLab is operated by the Association of Universities for Research in Astronomy (AURA), Inc. under a cooperative agreement with the U.S. National Science Foundation.
    
The Pan-STARRS1 Surveys (PS1) and the PS1 public science archive have been made possible through contributions by the Institute for Astronomy, the University of Hawaii, the Pan-STARRS Project Office, the Max-Planck Society and its participating institutes, the Max Planck Institute for Astronomy, Heidelberg and the Max Planck Institute for Extraterrestrial Physics, Garching, The Johns Hopkins University, Durham University, the University of Edinburgh, the Queen's University Belfast, the Harvard-Smithsonian Center for Astrophysics, the Las Cumbres Observatory Global Telescope Network Incorporated, the National Central University of Taiwan, the Space Telescope Science Institute, the National Aeronautics and Space Administration under Grant No. NNX08AR22G issued through the Planetary Science Division of the NASA Science Mission Directorate, the National Science Foundation Grant No.\ AST-1238877, the University of Maryland, Eotvos Lorand University (ELTE), the Los Alamos National Laboratory, and the Gordon and Betty Moore Foundation.

This research used the NASA/IPAC Extragalactic Database (NED),
which is operated by the Jet Propulsion Laboratory, California Institute of Technology,
under contract with the National Aeronautics and Space Administration.

This research made use of Montage, which is funded by the National Science Foundation under Grant Number ACI-1440620, and was previously funded by the National Aeronautics and Space Administration's Earth Science Technology Office, Computation Technologies Project, under Cooperative Agreement Number NCC5-626 between NASA and the California Institute of Technology.

\end{footnotesize}

\facilities{HST (WFC3, ACS), JWST (NIRCam), PS1, Blanco (DECam)}





\bibliographystyle{aasjournal}
\bibliography{jensen}{}

\begin{thebibliography}{}
\expandafter\ifx\csname natexlab\endcsname\relax\def\natexlab#1{#1}\fi
\providecommand{\url}[1]{\href{#1}{#1}}
\providecommand{\dodoi}[1]{doi:~\href{http://doi.org/#1}{\nolinkurl{#1}}}
\providecommand{\doeprint}[1]{\href{http://ascl.net/#1}{\nolinkurl{http://ascl.net/#1}}}
\providecommand{\doarXiv}[1]{\href{https://arxiv.org/abs/#1}{\nolinkurl{https://arxiv.org/abs/#1}}}

\bibitem[{{Anand} {et~al.}(2024{\natexlab{a}}){Anand}, {Riess}, {Yuan},
  {Beaton}, {Casertano}, {Li}, {Makarov}, {Makarova}, {Tully}, {Anderson},
  {Breuval}, {Dolphin}, {Karachentsev}, {Macri}, \& {Scolnic}}]{anand2024a}
{Anand}, G.~S., {Riess}, A.~G., {Yuan}, W., {et~al.} 2024{\natexlab{a}}, \apj,
  966, 89, \dodoi{10.3847/1538-4357/ad2e0a}

\bibitem[{{Anand} {et~al.}(2024{\natexlab{b}}){Anand}, {Tully}, {Cohen},
  {Makarov}, {Makarova}, {Jensen}, {Blakeslee}, {Cantiello}, {Kourkchi}, \&
  {Raimondo}}]{anand2024b}
{Anand}, G.~S., {Tully}, R.~B., {Cohen}, Y., {et~al.} 2024{\natexlab{b}}, \apj,
  973, 83, \dodoi{10.3847/1538-4357/ad64c7}

\bibitem[{{Anand} {et~al.}(2025){Anand}, {Tully}, {Cohen}, {Shaya}, {Makarov},
  {Makarova}, {Chazov}, {Blakeslee}, {Cantiello}, {Jensen}, {Kourkchi}, \&
  {Raimondo}}]{anand2024c}
---. 2025, arXiv e-prints, arXiv:2408.16810, \dodoi{10.48550/arXiv.2408.16810}

\bibitem[{{Bhardwaj} {et~al.}(2024){Bhardwaj}, {Ripepi}, {Testa}, {Molinaro},
  {Marconi}, {De Somma}, {Trentin}, {Musella}, {Storm}, {Sicignano}, \&
  {Catanzaro}}]{Bhardwaj2024}
{Bhardwaj}, A., {Ripepi}, V., {Testa}, V., {et~al.} 2024, \aap, 683, A234,
  \dodoi{10.1051/0004-6361/202348140}

\bibitem[{{Blakeslee} {et~al.}(2021){Blakeslee}, {Jensen}, {Ma}, {Milne}, \&
  {Greene}}]{blakeslee2021}
{Blakeslee}, J.~P., {Jensen}, J.~B., {Ma}, C.-P., {Milne}, P.~A., \& {Greene},
  J.~E. 2021, \apj, 911, 65, \dodoi{10.3847/1538-4357/abe86a}

\bibitem[{{Blakeslee} {et~al.}(2002){Blakeslee}, {Lucey}, {Tonry}, {Hudson},
  {Narayanan}, \& {Barris}}]{blakeslee2002}
{Blakeslee}, J.~P., {Lucey}, J.~R., {Tonry}, J.~L., {et~al.} 2002, \mnras, 330,
  443, \dodoi{10.1046/j.1365-8711.2002.05080.x}

\bibitem[{{Blakeslee} {et~al.}(2009){Blakeslee}, {Jord{\'a}n}, {Mei},
  {C{\^o}t{\'e}}, {Ferrarese}, {Infante}, {Peng}, {Tonry}, \&
  {West}}]{blakeslee2009}
{Blakeslee}, J.~P., {Jord{\'a}n}, A., {Mei}, S., {et~al.} 2009, \apj, 694, 556,
  \dodoi{10.1088/0004-637X/694/1/556}

\bibitem[{{Blakeslee} {et~al.}(2010){Blakeslee}, {Cantiello}, {Mei},
  {C{\^o}t{\'e}}, {Barber DeGraaff}, {Ferrarese}, {Jord{\'a}n}, {Peng},
  {Tonry}, \& {Worthey}}]{blakeslee2010}
{Blakeslee}, J.~P., {Cantiello}, M., {Mei}, S., {et~al.} 2010, \apj, 724, 657,
  \dodoi{10.1088/0004-637X/724/1/657}

\bibitem[{{Brout} \& {Riess}(2023)}]{Brout2023}
{Brout}, D., \& {Riess}, A. 2023, arXiv e-prints, arXiv:2311.08253,
  \dodoi{10.48550/arXiv.2311.08253}

\bibitem[{{Cantiello} {et~al.}(2018){Cantiello}, {Jensen}, {Blakeslee},
  {Berger}, {Levan}, {Tanvir}, {Raimondo}, {Brocato}, {Alexander}, {Blanchard},
  {Branchesi}, {Cano}, {Chornock}, {Covino}, {Cowperthwaite}, {D'Avanzo},
  {Eftekhari}, {Fong}, {Fruchter}, {Grado}, {Hjorth}, {Holz}, {Lyman},
  {Mandel}, {Margutti}, {Nicholl}, {Villar}, \& {Williams}}]{cantiello2018}
{Cantiello}, M., {Jensen}, J.~B., {Blakeslee}, J.~P., {et~al.} 2018, \apjl,
  854, L31, \dodoi{10.3847/2041-8213/aaad64}

\bibitem[{{Cantiello} {et~al.}(2024){Cantiello}, {Blakeslee}, {Ferrarese},
  {C{\^o}t{\'e}}, {Raimondo}, {Cuillandre}, {Durrell}, {Gwyn}, {Hazra}, {Peng},
  {Roediger}, {S{\'a}nchez-Janssen}, \& {Kurzner}}]{Cantiello2024}
{Cantiello}, M., {Blakeslee}, J.~P., {Ferrarese}, L., {et~al.} 2024, \apj, 966,
  145, \dodoi{10.3847/1538-4357/ad3453}

\bibitem[{{Carrick} {et~al.}(2015){Carrick}, {Turnbull}, {Lavaux}, \&
  {Hudson}}]{2M++2015}
{Carrick}, J., {Turnbull}, S.~J., {Lavaux}, G., \& {Hudson}, M.~J. 2015,
  \mnras, 450, 317, \dodoi{10.1093/mnras/stv547}

\bibitem[{Chambers {et~al.}(2019)Chambers, Magnier, Metcalfe, Flewelling,
  Huber, Waters, Denneau, Draper, Farrow, Finkbeiner, Holmberg, Koppenhoefer,
  Price, Rest, Saglia, Schlafly, Smartt, Sweeney, Wainscoat, Burgett, Chastel,
  Grav, Heasley, Hodapp, Jedicke, Kaiser, Kudritzki, Luppino, Lupton, Monet,
  Morgan, Onaka, Shiao, Stubbs, Tonry, White, Bañados, Bell, Bender, Bernard,
  Boegner, Boffi, Botticella, Calamida, Casertano, Chen, Chen, Cole, Deacon,
  Frenk, Fitzsimmons, Gezari, Gibbs, Goessl, Goggia, Gourgue, Goldman, Grant,
  Grebel, Hambly, Hasinger, Heavens, Heckman, Henderson, Henning, Holman, Hopp,
  Ip, Isani, Jackson, Keyes, Koekemoer, Kotak, Le, Liska, Long, Lucey, Liu,
  Martin, Masci, McLean, Mindel, Misra, Morganson, Murphy, Obaika, Narayan,
  Nieto-Santisteban, Norberg, Peacock, Pier, Postman, Primak, Rae, Rai, Riess,
  Riffeser, Rix, Röser, Russel, Rutz, Schilbach, Schultz, Scolnic, Strolger,
  Szalay, Seitz, Small, Smith, Soderblom, Taylor, Thomson, Taylor, Thakar,
  Thiel, Thilker, Unger, Urata, Valenti, Wagner, Walder, Walter, Watters,
  Werner, Wood-Vasey, \& Wyse}]{chambers2019panstarrs1surveys}
Chambers, K.~C., Magnier, E.~A., Metcalfe, N., {et~al.} 2019, The Pan-STARRS1
  Surveys.
\newblock \doarXiv{1612.05560}

\bibitem[{{Dey} {et~al.}(2019){Dey}, {Schlegel}, {Lang}, {Blum}, {Burleigh},
  {Fan}, {Findlay}, {Finkbeiner}, {Herrera}, {Juneau}, {Landriau}, {Levi},
  {McGreer}, {Meisner}, {Myers}, {Moustakas}, {Nugent}, {Patej}, {Schlafly},
  {Walker}, {Valdes}, {Weaver}, {Y{\`e}che}, {Zou}, {Zhou}, {Abareshi},
  {Abbott}, {Abolfathi}, {Aguilera}, {Alam}, {Allen}, {Alvarez}, {Annis},
  {Ansarinejad}, {Aubert}, {Beechert}, {Bell}, {BenZvi}, {Beutler}, {Bielby},
  {Bolton}, {Brice{\~n}o}, {Buckley-Geer}, {Butler}, {Calamida}, {Carlberg},
  {Carter}, {Casas}, {Castander}, {Choi}, {Comparat}, {Cukanovaite}, {Delubac},
  {DeVries}, {Dey}, {Dhungana}, {Dickinson}, {Ding}, {Donaldson}, {Duan},
  {Duckworth}, {Eftekharzadeh}, {Eisenstein}, {Etourneau}, {Fagrelius},
  {Farihi}, {Fitzpatrick}, {Font-Ribera}, {Fulmer}, {G{\"a}nsicke},
  {Gaztanaga}, {George}, {Gerdes}, {Gontcho}, {Gorgoni}, {Green}, {Guy},
  {Harmer}, {Hernandez}, {Honscheid}, {Huang}, {James}, {Jannuzi}, {Jiang},
  {Joyce}, {Karcher}, {Karkar}, {Kehoe}, {Kneib}, {Kueter-Young}, {Lan},
  {Lauer}, {Le Guillou}, {Le Van Suu}, {Lee}, {Lesser}, {Perreault Levasseur},
  {Li}, {Mann}, {Marshall}, {Mart{\'\i}nez-V{\'a}zquez}, {Martini}, {du Mas des
  Bourboux}, {McManus}, {Meier}, {M{\'e}nard}, {Metcalfe},
  {Mu{\~n}oz-Guti{\'e}rrez}, {Najita}, {Napier}, {Narayan}, {Newman}, {Nie},
  {Nord}, {Norman}, {Olsen}, {Paat}, {Palanque-Delabrouille}, {Peng},
  {Poppett}, {Poremba}, {Prakash}, {Rabinowitz}, {Raichoor}, {Rezaie},
  {Robertson}, {Roe}, {Ross}, {Ross}, {Rudnick}, {Safonova}, {Saha},
  {S{\'a}nchez}, {Savary}, {Schweiker}, {Scott}, {Seo}, {Shan}, {Silva},
  {Slepian}, {Soto}, {Sprayberry}, {Staten}, {Stillman}, {Stupak}, {Summers},
  {Sien Tie}, {Tirado}, {Vargas-Maga{\~n}a}, {Vivas}, {Wechsler}, {Williams},
  {Yang}, {Yang}, {Yapici}, {Zaritsky}, {Zenteno}, {Zhang}, {Zhang}, {Zhou}, \&
  {Zhou}}]{decamphotometry2019}
{Dey}, A., {Schlegel}, D.~J., {Lang}, D., {et~al.} 2019, \aj, 157, 168,
  \dodoi{10.3847/1538-3881/ab089d}

\bibitem[{{Di Valentino} {et~al.}(2021){Di Valentino}, {Mena}, {Pan},
  {Visinelli}, {Yang}, {Melchiorri}, {Mota}, {Riess}, \&
  {Silk}}]{divalentino2021}
{Di Valentino}, E., {Mena}, O., {Pan}, S., {et~al.} 2021, Classical and Quantum
  Gravity, 38, 153001, \dodoi{10.1088/1361-6382/ac086d}

\bibitem[{{Di Valentino} {et~al.}(2025){Di Valentino}, {Levi Said}, {Riess},
  {Pollo}, {Poulin}, {G{\'o}mez-Valent}, {Weltman}, {Palmese}, {Huang}, {van de
  Bruck}, {Shekhar Saraf}, {Kuo}, {Uhlemann}, {Grand{\'o}n}, {Paz}, {Eckert},
  {Teixeira}, {Saridakis}, {Colg{\'a}in}, {Beutler}, {Niedermann}, {Bajardi},
  {Barenboim}, {Gubitosi}, {Musella}, {Banik}, {Szapudi}, {Singal}, {Haro
  Cases}, {Chluba}, {Torrado}, {Mifsud}, {Jedamzik}, {Said}, {Dialektopoulos},
  {Herold}, {Perivolaropoulos}, {Zu}, {Galbany}, {Breuval}, {Visinelli},
  {Escamilla}, {Anchordoqui}, {Sheikh-Jabbari}, {Lembo}, {Dainotti},
  {Vincenzi}, {Asgari}, {Gerbino}, {Forconi}, {Cantiello}, {Moresco},
  {Benetti}, {Sch{\"o}neberg}, {Akarsu}, {Nunes}, {Bernardo}, {Ch{\'a}vez},
  {Anderson}, {Watkins}, {Capozziello}, {Li}, {Vagnozzi}, {Pan}, {Treu},
  {Irsic}, {Handley}, {Giar{\`e}}, {Murakami}, {Poudou}, {Heavens}, {Kogut},
  {Domi}, {{\L}ukasz Lenart}, {Melchiorri}, {Vadal{\`a}}, {Amon}, {Bonilla},
  {Reeves}, {Zhuk}, {Bonanno}, {{\"O}vg{\"u}n}, {Pisani}, {Talebian}, {Abebe},
  {Aboubrahim}, {Gonz{\'a}lez Mor{\'a}n}, {Kov{\'a}cs}, {Papatriantafyllou},
  {Liddle}, {Paliathanasis}, {Borowiec}, {Yadav}, {Yadav}, {Sen}, {Mini Latha},
  {Davis}, {Shajib}, {Walters}, {Idicherian Lonappan}, {Chudaykin},
  {Capodagli}, {da Silva}, {De Felice}, {Racioppi}, {Soler Oficial}, {Montiel},
  {Favale}, {Bernui}, {Velasco}, {Heinesen}, {Bakopoulos}, {Chatzistavrakidis},
  {Khanpour}, {Sathyaprakash}, {Zgirski}, {L'Huillier}, {Famaey}, {Jain},
  {Marek}, {Zhang}, {Karmakar}, {Dragovich}, {Thomas}, {Correa}, {Boiza},
  {Marques}, {Escamilla-Rivera}, {Tzerefos}, {Zhang}, {De Leo}, {Pfeifer},
  {Lee}, {Venter}, {Gomes}, {Roque De bom}, {Moreno-Pulido}, {Iosifidis},
  {Grin}, {Blixt}, {Scolnic}, {Oriti}, {Dobrycheva}, {Bettoni}, {Benisty},
  {Fern{\'a}ndez-Arenas}, {Wiltshire}, {Sanchez Cid}, {Tamayo}, {Valls-Gabaud},
  {Pedrotti}, {Wang}, {Staicova}, {Totolou}, {Rubiera-Garcia}, {Milakovi{\'c}},
  {Pesce}, {Sluse}, {Borka}, {Yusofi}, {Giusarma}, {Terlevich}, {Tomasetti},
  {Vagenas}, {Fazzari}, {Ferreira}, {Barakovic}, {Dimastrogiovanni}, {Brinch
  Holm}, {Mottola}, {{\"O}z{\"u}lker}, {Specogna}, {Brocato}, {Jensko},
  {Antonette Enriquez}, {Bhatia}, {Bresolin}, {Avila}, {Bouch{\`e}},
  {Bombacigno}, {Anagnostopoulos}, {Pace}, {Sorrenti}, {Lobo}, {Courbin},
  {Hansen}, {Sloan}, {Farrugia}, {Lynch}, {Garcia-Arroyo}, {Raimondo},
  {Lambiase}, {Anand}, {Poulot}, {Leon}, {Kouniatalis}, {Nardini},
  {Cs{\"o}rnyei}, {Galloni}, \& {Bargiacchi}}]{CosmoVerse2025}
{Di Valentino}, E., {Levi Said}, J., {Riess}, A., {et~al.} 2025, arXiv
  e-prints, arXiv:2504.01669, \dodoi{10.48550/arXiv.2504.01669}

\bibitem[{{Foreman-Mackey} {et~al.}(2013){Foreman-Mackey}, {Hogg}, {Lang}, \&
  {Goodman}}]{emcee2013}
{Foreman-Mackey}, D., {Hogg}, D.~W., {Lang}, D., \& {Goodman}, J. 2013, \pasp,
  125, 306, \dodoi{10.1086/670067}

\bibitem[{{Freedman} {et~al.}(2024){Freedman}, {Madore}, {Jang}, {Hoyt}, {Lee},
  \& {Owens}}]{freedman2024}
{Freedman}, W.~L., {Madore}, B.~F., {Jang}, I.~S., {et~al.} 2024, arXiv
  e-prints, arXiv:2408.06153, \dodoi{10.48550/arXiv.2408.06153}

\bibitem[{{Gaia Collaboration} {et~al.}(2016){Gaia Collaboration}, {Prusti,
  T.}, {de Bruijne, J. H. J.}, {Brown, A. G. A.}, {Vallenari, A.}, {Babusiaux,
  C.}, {Bailer-Jones, C. A. L.}, {Bastian, U.}, {Biermann, M.}, {Evans, D. W.},
  {Eyer, L.}, {Jansen, F.}, {Jordi, C.}, {Klioner, S. A.}, {Lammers, U.},
  {Lindegren, L.}, {Luri, X.}, {Mignard, F.}, {Milligan, D. J.}, {Panem, C.},
  {Poinsignon, V.}, {Pourbaix, D.}, {Randich, S.}, {Sarri, G.}, {Sartoretti,
  P.}, {Siddiqui, H. I.}, {Soubiran, C.}, {Valette, V.}, {van Leeuwen, F.},
  {Walton, N. A.}, {Aerts, C.}, {Arenou, F.}, {Cropper, M.}, {Drimmel, R.},
  {Høg, E.}, {Katz, D.}, {Lattanzi, M. G.}, {O’Mullane, W.}, {Grebel, E.
  K.}, {Holland, A. D.}, {Huc, C.}, {Passot, X.}, {Bramante, L.}, {Cacciari,
  C.}, {Castañeda, J.}, {Chaoul, L.}, {Cheek, N.}, {De Angeli, F.},
  {Fabricius, C.}, {Guerra, R.}, {Hernández, J.}, {Jean-Antoine-Piccolo, A.},
  {Masana, E.}, {Messineo, R.}, {Mowlavi, N.}, {Nienartowicz, K.},
  {Ordóñez-Blanco, D.}, {Panuzzo, P.}, {Portell, J.}, {Richards, P. J.},
  {Riello, M.}, {Seabroke, G. M.}, {Tanga, P.}, {Thévenin, F.}, {Torra, J.},
  {Els, S. G.}, {Gracia-Abril, G.}, {Comoretto, G.}, {Garcia-Reinaldos, M.},
  {Lock, T.}, {Mercier, E.}, {Altmann, M.}, {Andrae, R.}, {Astraatmadja, T.
  L.}, {Bellas-Velidis, I.}, {Benson, K.}, {Berthier, J.}, {Blomme, R.},
  {Busso, G.}, {Carry, B.}, {Cellino, A.}, {Clementini, G.}, {Cowell, S.},
  {Creevey, O.}, {Cuypers, J.}, {Davidson, M.}, {De Ridder, J.}, {de Torres,
  A.}, {Delchambre, L.}, {Dell’Oro, A.}, {Ducourant, C.}, {Frémat, Y.},
  {García-Torres, M.}, {Gosset, E.}, {Halbwachs, J.-L.}, {Hambly, N. C.},
  {Harrison, D. L.}, {Hauser, M.}, {Hestroffer, D.}, {Hodgkin, S. T.}, {Huckle,
  H. E.}, {Hutton, A.}, {Jasniewicz, G.}, {Jordan, S.}, {Kontizas, M.}, {Korn,
  A. J.}, {Lanzafame, A. C.}, {Manteiga, M.}, {Moitinho, A.}, {Muinonen, K.},
  {Osinde, J.}, {Pancino, E.}, {Pauwels, T.}, {Petit, J.-M.}, {Recio-Blanco,
  A.}, {Robin, A. C.}, {Sarro, L. M.}, {Siopis, C.}, {Smith, M.}, {Smith, K.
  W.}, {Sozzetti, A.}, {Thuillot, W.}, {van Reeven, W.}, {Viala, Y.}, {Abbas,
  U.}, {Abreu Aramburu, A.}, {Accart, S.}, {Aguado, J. J.}, {Allan, P. M.},
  {Allasia, W.}, {Altavilla, G.}, {Álvarez, M. A.}, {Alves, J.}, {Anderson, R.
  I.}, {Andrei, A. H.}, {Anglada Varela, E.}, {Antiche, E.}, {Antoja, T.},
  {Antón, S.}, {Arcay, B.}, {Atzei, A.}, {Ayache, L.}, {Bach, N.}, {Baker, S.
  G.}, {Balaguer-Núñez, L.}, {Barache, C.}, {Barata, C.}, {Barbier, A.},
  {Barblan, F.}, {Baroni, M.}, {Barrado y Navascués, D.}, {Barros, M.},
  {Barstow, M. A.}, {Becciani, U.}, {Bellazzini, M.}, {Bellei, G.}, {Bello
  García, A.}, {Belokurov, V.}, {Bendjoya, P.}, {Berihuete, A.}, {Bianchi,
  L.}, {Bienaymé, O.}, {Billebaud, F.}, {Blagorodnova, N.}, {Blanco-Cuaresma,
  S.}, {Boch, T.}, {Bombrun, A.}, {Borrachero, R.}, {Bouquillon, S.}, {Bourda,
  G.}, {Bouy, H.}, {Bragaglia, A.}, {Breddels, M. A.}, {Brouillet, N.},
  {Brüsemeister, T.}, {Bucciarelli, B.}, {Budnik, F.}, {Burgess, P.}, {Burgon,
  R.}, {Burlacu, A.}, {Busonero, D.}, {Buzzi, R.}, {Caffau, E.}, {Cambras, J.},
  {Campbell, H.}, {Cancelliere, R.}, {Cantat-Gaudin, T.}, {Carlucci, T.},
  {Carrasco, J. M.}, {Castellani, M.}, {Charlot, P.}, {Charnas, J.}, {Charvet,
  P.}, {Chassat, F.}, {Chiavassa, A.}, {Clotet, M.}, {Cocozza, G.}, {Collins,
  R. S.}, {Collins, P.}, {Costigan, G.}, {Crifo, F.}, {Cross, N. J. G.},
  {Crosta, M.}, {Crowley, C.}, {Dafonte, C.}, {Damerdji, Y.}, {Dapergolas, A.},
  {David, P.}, {David, M.}, {De Cat, P.}, {de Felice, F.}, {de Laverny, P.},
  {De Luise, F.}, {De March, R.}, {de Martino, D.}, {de Souza, R.},
  {Debosscher, J.}, {del Pozo, E.}, {Delbo, M.}, {Delgado, A.}, {Delgado, H.
  E.}, {di Marco, F.}, {Di Matteo, P.}, {Diakite, S.}, {Distefano, E.},
  {Dolding, C.}, {Dos Anjos, S.}, {Drazinos, P.}, {Durán, J.}, {Dzigan, Y.},
  {Ecale, E.}, {Edvardsson, B.}, {Enke, H.}, {Erdmann, M.}, {Escolar, D.},
  {Espina, M.}, {Evans, N. W.}, {Eynard Bontemps, G.}, {Fabre, C.}, {Fabrizio,
  M.}, {Faigler, S.}, {Falcão, A. J.}, {Farràs Casas, M.}, {Faye, F.},
  {Federici, L.}, {Fedorets, G.}, {Fernández-Hernández, J.}, {Fernique, P.},
  {Fienga, A.}, {Figueras, F.}, {Filippi, F.}, {Findeisen, K.}, {Fonti, A.},
  {Fouesneau, M.}, {Fraile, E.}, {Fraser, M.}, {Fuchs, J.}, {Furnell, R.},
  {Gai, M.}, {Galleti, S.}, {Galluccio, L.}, {Garabato, D.}, {García-Sedano,
  F.}, {Garé, P.}, {Garofalo, A.}, {Garralda, N.}, {Gavras, P.}, {Gerssen,
  J.}, {Geyer, R.}, {Gilmore, G.}, {Girona, S.}, {Giuffrida, G.}, {Gomes, M.},
  {González-Marcos, A.}, {González-Núñez, J.}, {González-Vidal, J. J.},
  {Granvik, M.}, {Guerrier, A.}, {Guillout, P.}, {Guiraud, J.}, {Gúrpide, A.},
  {Gutiérrez-Sánchez, R.}, {Guy, L. P.}, {Haigron, R.}, {Hatzidimitriou, D.},
  {Haywood, M.}, {Heiter, U.}, {Helmi, A.}, {Hobbs, D.}, {Hofmann, W.}, {Holl,
  B.}, {Holland, G.}, {Hunt, J. A. S.}, {Hypki, A.}, {Icardi, V.}, {Irwin, M.},
  {Jevardat de Fombelle, G.}, {Jofré, P.}, {Jonker, P. G.}, {Jorissen, A.},
  {Julbe, F.}, {Karampelas, A.}, {Kochoska, A.}, {Kohley, R.}, {Kolenberg, K.},
  {Kontizas, E.}, {Koposov, S. E.}, {Kordopatis, G.}, {Koubsky, P.},
  {Kowalczyk, A.}, {Krone-Martins, A.}, {Kudryashova, M.}, {Kull, I.},
  {Bachchan, R. K.}, {Lacoste-Seris, F.}, {Lanza, A. F.}, {Lavigne, J.-B.}, {Le
  Poncin-Lafitte, C.}, {Lebreton, Y.}, {Lebzelter, T.}, {Leccia, S.}, {Leclerc,
  N.}, {Lecoeur-Taibi, I.}, {Lemaitre, V.}, {Lenhardt, H.}, {Leroux, F.},
  {Liao, S.}, {Licata, E.}, {Lindstrøm, H. E. P.}, {Lister, T. A.}, {Livanou,
  E.}, {Lobel, A.}, {Löffler, W.}, {López, M.}, {Lopez-Lozano, A.}, {Lorenz,
  D.}, {Loureiro, T.}, {MacDonald, I.}, {Magalhães Fernandes, T.}, {Managau,
  S.}, {Mann, R. G.}, {Mantelet, G.}, {Marchal, O.}, {Marchant, J. M.},
  {Marconi, M.}, {Marie, J.}, {Marinoni, S.}, {Marrese, P. M.}, {Marschalkó,
  G.}, {Marshall, D. J.}, {Martín-Fleitas, J. M.}, {Martino, M.}, {Mary, N.},
  {Matijevič, G.}, {Mazeh, T.}, {McMillan, P. J.}, {Messina, S.}, {Mestre,
  A.}, {Michalik, D.}, {Millar, N. R.}, {Miranda, B. M. H.}, {Molina, D.},
  {Molinaro, R.}, {Molinaro, M.}, {Molnár, L.}, {Moniez, M.}, {Montegriffo,
  P.}, {Monteiro, D.}, {Mor, R.}, {Mora, A.}, {Morbidelli, R.}, {Morel, T.},
  {Morgenthaler, S.}, {Morley, T.}, {Morris, D.}, {Mulone, A. F.}, {Muraveva,
  T.}, {Musella, I.}, {Narbonne, J.}, {Nelemans, G.}, {Nicastro, L.}, {Noval,
  L.}, {Ordénovic, C.}, {Ordieres-Meré, J.}, {Osborne, P.}, {Pagani, C.},
  {Pagano, I.}, {Pailler, F.}, {Palacin, H.}, {Palaversa, L.}, {Parsons, P.},
  {Paulsen, T.}, {Pecoraro, M.}, {Pedrosa, R.}, {Pentikäinen, H.}, {Pereira,
  J.}, {Pichon, B.}, {Piersimoni, A. M.}, {Pineau, F.-X.}, {Plachy, E.}, {Plum,
  G.}, {Poujoulet, E.}, {Prša, A.}, {Pulone, L.}, {Ragaini, S.}, {Rago, S.},
  {Rambaux, N.}, {Ramos-Lerate, M.}, {Ranalli, P.}, {Rauw, G.}, {Read, A.},
  {Regibo, S.}, {Renk, F.}, {Reylé, C.}, {Ribeiro, R. A.}, {Rimoldini, L.},
  {Ripepi, V.}, {Riva, A.}, {Rixon, G.}, {Roelens, M.}, {Romero-Gómez, M.},
  {Rowell, N.}, {Royer, F.}, {Rudolph, A.}, {Ruiz-Dern, L.}, {Sadowski, G.},
  {Sagristà Sellés, T.}, {Sahlmann, J.}, {Salgado, J.}, {Salguero, E.},
  {Sarasso, M.}, {Savietto, H.}, {Schnorhk, A.}, {Schultheis, M.}, {Sciacca,
  E.}, {Segol, M.}, {Segovia, J. C.}, {Segransan, D.}, {Serpell, E.}, {Shih,
  I-C.}, {Smareglia, R.}, {Smart, R. L.}, {Smith, C.}, {Solano, E.}, {Solitro,
  F.}, {Sordo, R.}, {Soria Nieto, S.}, {Souchay, J.}, {Spagna, A.}, {Spoto,
  F.}, {Stampa, U.}, {Steele, I. A.}, {Steidelmüller, H.}, {Stephenson, C.
  A.}, {Stoev, H.}, {Suess, F. F.}, {Süveges, M.}, {Surdej, J.}, {Szabados,
  L.}, {Szegedi-Elek, E.}, {Tapiador, D.}, {Taris, F.}, {Tauran, G.}, {Taylor,
  M. B.}, {Teixeira, R.}, {Terrett, D.}, {Tingley, B.}, {Trager, S. C.},
  {Turon, C.}, {Ulla, A.}, {Utrilla, E.}, {Valentini, G.}, {van Elteren, A.},
  {Van Hemelryck, E.}, {van Leeuwen, M.}, {Varadi, M.}, {Vecchiato, A.},
  {Veljanoski, J.}, {Via, T.}, {Vicente, D.}, {Vogt, S.}, {Voss, H.}, {Votruba,
  V.}, {Voutsinas, S.}, {Walmsley, G.}, {Weiler, M.}, {Weingrill, K.}, {Werner,
  D.}, {Wevers, T.}, {Whitehead, G.}, {Wyrzykowski, Ł.}, {Yoldas, A.},
  {Žerjal, M.}, {Zucker, S.}, {Zurbach, C.}, {Zwitter, T.}, {Alecu, A.},
  {Allen, M.}, {Allende Prieto, C.}, {Amorim, A.}, {Anglada-Escudé, G.},
  {Arsenijevic, V.}, {Azaz, S.}, {Balm, P.}, {Beck, M.}, {Bernstein, H.-H.},
  {Bigot, L.}, {Bijaoui, A.}, {Blasco, C.}, {Bonfigli, M.}, {Bono, G.},
  {Boudreault, S.}, {Bressan, A.}, {Brown, S.}, {Brunet, P.-M.}, {Bunclark,
  P.}, {Buonanno, R.}, {Butkevich, A. G.}, {Carret, C.}, {Carrion, C.},
  {Chemin, L.}, {Chéreau, F.}, {Corcione, L.}, {Darmigny, E.}, {de Boer, K.
  S.}, {de Teodoro, P.}, {de Zeeuw, P. T.}, {Delle Luche, C.}, {Domingues, C.
  D.}, {Dubath, P.}, {Fodor, F.}, {Frézouls, B.}, {Fries, A.}, {Fustes, D.},
  {Fyfe, D.}, {Gallardo, E.}, {Gallegos, J.}, {Gardiol, D.}, {Gebran, M.},
  {Gomboc, A.}, {Gómez, A.}, {Grux, E.}, {Gueguen, A.}, {Heyrovsky, A.},
  {Hoar, J.}, {Iannicola, G.}, {Isasi Parache, Y.}, {Janotto, A.-M.}, {Joliet,
  E.}, {Jonckheere, A.}, {Keil, R.}, {Kim, D.-W.}, {Klagyivik, P.}, {Klar, J.},
  {Knude, J.}, {Kochukhov, O.}, {Kolka, I.}, {Kos, J.}, {Kutka, A.}, {Lainey,
  V.}, {LeBouquin, D.}, {Liu, C.}, {Loreggia, D.}, {Makarov, V. V.},
  {Marseille, M. G.}, {Martayan, C.}, {Martinez-Rubi, O.}, {Massart, B.},
  {Meynadier, F.}, {Mignot, S.}, {Munari, U.}, {Nguyen, A.-T.}, {Nordlander,
  T.}, {Ocvirk, P.}, {O’Flaherty, K. S.}, {Olias Sanz, A.}, {Ortiz, P.},
  {Osorio, J.}, {Oszkiewicz, D.}, {Ouzounis, A.}, {Palmer, M.}, {Park, P.},
  {Pasquato, E.}, {Peltzer, C.}, {Peralta, J.}, {Péturaud, F.}, {Pieniluoma,
  T.}, {Pigozzi, E.}, {Poels, J.}, {Prat, G.}, {Prod’homme, T.}, {Raison,
  F.}, {Rebordao, J. M.}, {Risquez, D.}, {Rocca-Volmerange, B.}, {Rosen, S.},
  {Ruiz-Fuertes, M. I.}, {Russo, F.}, {Sembay, S.}, {Serraller Vizcaino, I.},
  {Short, A.}, {Siebert, A.}, {Silva, H.}, {Sinachopoulos, D.}, {Slezak, E.},
  {Soffel, M.}, {Sosnowska, D.}, {Straižys, V.}, {ter Linden, M.}, {Terrell,
  D.}, {Theil, S.}, {Tiede, C.}, {Troisi, L.}, {Tsalmantza, P.}, {Tur, D.},
  {Vaccari, M.}, {Vachier, F.}, {Valles, P.}, {Van Hamme, W.}, {Veltz, L.},
  {Virtanen, J.}, {Wallut, J.-M.}, {Wichmann, R.}, {Wilkinson, M. I.},
  {Ziaeepour, H.}, \& {Zschocke, S.}}]{Gaia2016}
{Gaia Collaboration}, {Prusti, T.}, {de Bruijne, J. H. J.}, {et~al.} 2016,
  A\&A, 595, A1, \dodoi{10.1051/0004-6361/201629272}

\bibitem[{{Gaia Collaboration} {et~al.}(2017){Gaia Collaboration},
  {Clementini}, {Eyer}, {Ripepi}, {Marconi}, {Muraveva}, {Garofalo}, {Sarro},
  {Palmer}, {Luri}, {Molinaro}, {Rimoldini}, {Szabados}, {Musella}, {Anderson},
  {Prusti}, {de Bruijne}, {Brown}, {Vallenari}, {Babusiaux}, {Bailer-Jones},
  {Bastian}, {Biermann}, {Evans}, {Jansen}, {Jordi}, {Klioner}, {Lammers},
  {Lindegren}, {Mignard}, {Panem}, {Pourbaix}, {Randich}, {Sartoretti},
  {Siddiqui}, {Soubiran}, {Valette}, {van Leeuwen}, {Walton}, {Aerts},
  {Arenou}, {Cropper}, {Drimmel}, {H{\o}g}, {Katz}, {Lattanzi}, {O'Mullane},
  {Grebel}, {Holland}, {Huc}, {Passot}, {Perryman}, {Bramante}, {Cacciari},
  {Casta{\~n}eda}, {Chaoul}, {Cheek}, {De Angeli}, {Fabricius}, {Guerra},
  {Hern{\'a}ndez}, {Jean-Antoine-Piccolo}, {Masana}, {Messineo}, {Mowlavi},
  {Nienartowicz}, {Ord{\'o}{\~n}ez-Blanco}, {Panuzzo}, {Portell}, {Richards},
  {Riello}, {Seabroke}, {Tanga}, {Th{\'e}venin}, {Torra}, {Els},
  {Gracia-Abril}, {Comoretto}, {Garcia-Reinaldos}, {Lock}, {Mercier},
  {Altmann}, {Andrae}, {Astraatmadja}, {Bellas-Velidis}, {Benson}, {Berthier},
  {Blomme}, {Busso}, {Carry}, {Cellino}, {Cowell}, {Creevey}, {Cuypers},
  {Davidson}, {De Ridder}, {de Torres}, {Delchambre}, {Dell'Oro}, {Ducourant},
  {Fr{\'e}mat}, {Garc{\'\i}a-Torres}, {Gosset}, {Halbwachs}, {Hambly},
  {Harrison}, {Hauser}, {Hestroffer}, {Hodgkin}, {Huckle}, {Hutton},
  {Jasniewicz}, {Jordan}, {Kontizas}, {Korn}, {Lanzafame}, {Manteiga},
  {Moitinho}, {Muinonen}, {Osinde}, {Pancino}, {Pauwels}, {Petit},
  {Recio-Blanco}, {Robin}, {Siopis}, {Smith}, {Smith}, {Sozzetti}, {Thuillot},
  {van Reeven}, {Viala}, {Abbas}, {Abreu Aramburu}, {Accart}, {Aguado},
  {Allan}, {Allasia}, {Altavilla}, {{\'A}lvarez}, {Alves}, {Andrei}, {Anglada
  Varela}, {Antiche}, {Antoja}, {Ant{\'o}n}, {Arcay}, {Bach}, {Baker},
  {Balaguer-N{\'u}{\~n}ez}, {Barache}, {Barata}, {Barbier}, {Barblan}, {Barrado
  y Navascu{\'e}s}, {Barros}, {Barstow}, {Becciani}, {Bellazzini}, {Bello
  Garc{\'\i}a}, {Belokurov}, {Bendjoya}, {Berihuete}, {Bianchi},
  {Bienaym{\'e}}, {Billebaud}, {Blagorodnova}, {Blanco-Cuaresma}, {Boch},
  {Bombrun}, {Borrachero}, {Bouquillon}, {Bourda}, {Bragaglia}, {Breddels},
  {Brouillet}, {Br{\"u}semeister}, {Bucciarelli}, {Burgess}, {Burgon},
  {Burlacu}, {Busonero}, {Buzzi}, {Caffau}, {Cambras}, {Campbell},
  {Cancelliere}, {Cantat-Gaudin}, {Carlucci}, {Carrasco}, {Castellani},
  {Charlot}, {Charnas}, {Chiavassa}, {Clotet}, {Cocozza}, {Collins},
  {Costigan}, {Crifo}, \& {Cross}}]{Gaia2017}
{Gaia Collaboration}, {Clementini}, G., {Eyer}, L., {et~al.} 2017, \aap, 605,
  A79, \dodoi{10.1051/0004-6361/201629925}

\bibitem[{{Gaia Collaboration} {et~al.}(2018){Gaia Collaboration}, {Brown},
  {Vallenari}, {Prusti}, {de Bruijne}, {Babusiaux}, {Bailer-Jones}, {Biermann},
  {Evans}, {Eyer}, {Jansen}, {Jordi}, {Klioner}, {Lammers}, {Lindegren},
  {Luri}, {Mignard}, {Panem}, {Pourbaix}, {Randich}, {Sartoretti}, {Siddiqui},
  {Soubiran}, {van Leeuwen}, {Walton}, {Arenou}, {Bastian}, {Cropper},
  {Drimmel}, {Katz}, {Lattanzi}, {Bakker}, {Cacciari}, {Casta{\~n}eda},
  {Chaoul}, {Cheek}, {De Angeli}, {Fabricius}, {Guerra}, {Holl}, {Masana},
  {Messineo}, {Mowlavi}, {Nienartowicz}, {Panuzzo}, {Portell}, {Riello},
  {Seabroke}, {Tanga}, {Th{\'e}venin}, {Gracia-Abril}, {Comoretto},
  {Garcia-Reinaldos}, {Teyssier}, {Altmann}, {Andrae}, {Audard},
  {Bellas-Velidis}, {Benson}, {Berthier}, {Blomme}, {Burgess}, {Busso},
  {Carry}, {Cellino}, {Clementini}, {Clotet}, {Creevey}, {Davidson}, {De
  Ridder}, {Delchambre}, {Dell'Oro}, {Ducourant},
  {Fern{\'a}ndez-Hern{\'a}ndez}, {Fouesneau}, {Fr{\'e}mat}, {Galluccio},
  {Garc{\'\i}a-Torres}, {Gonz{\'a}lez-N{\'u}{\~n}ez}, {Gonz{\'a}lez-Vidal},
  {Gosset}, {Guy}, {Halbwachs}, {Hambly}, {Harrison}, {Hern{\'a}ndez},
  {Hestroffer}, {Hodgkin}, {Hutton}, {Jasniewicz}, {Jean-Antoine-Piccolo},
  {Jordan}, {Korn}, {Krone-Martins}, {Lanzafame}, {Lebzelter}, {L{\"o}ffler},
  {Manteiga}, {Marrese}, {Mart{\'\i}n-Fleitas}, {Moitinho}, {Mora}, {Muinonen},
  {Osinde}, {Pancino}, {Pauwels}, {Petit}, {Recio-Blanco}, {Richards},
  {Rimoldini}, {Robin}, {Sarro}, {Siopis}, {Smith}, {Sozzetti}, {S{\"u}veges},
  {Torra}, {van Reeven}, {Abbas}, {Abreu Aramburu}, {Accart}, {Aerts},
  {Altavilla}, {{\'A}lvarez}, {Alvarez}, {Alves}, {Anderson}, {Andrei},
  {Anglada Varela}, {Antiche}, {Antoja}, {Arcay}, {Astraatmadja}, {Bach},
  {Baker}, {Balaguer-N{\'u}{\~n}ez}, {Balm}, {Barache}, {Barata}, {Barbato},
  {Barblan}, {Barklem}, {Barrado}, {Barros}, {Barstow}, {Bartholom{\'e}
  Mu{\~n}oz}, {Bassilana}, {Becciani}, {Bellazzini}, {Berihuete}, {Bertone},
  {Bianchi}, {Bienaym{\'e}}, {Blanco-Cuaresma}, {Boch}, {Boeche}, {Bombrun},
  {Borrachero}, {Bossini}, {Bouquillon}, {Bourda}, {Bragaglia}, {Bramante},
  {Breddels}, {Bressan}, {Brouillet}, {Br{\"u}semeister}, {Brugaletta},
  {Bucciarelli}, {Burlacu}, {Busonero}, {Butkevich}, {Buzzi}, {Caffau},
  {Cancelliere}, {Cannizzaro}, {Cantat-Gaudin}, {Carballo}, {Carlucci},
  {Carrasco}, {Casamiquela}, {Castellani}, {Castro-Ginard}, {Charlot},
  {Chemin}, {Chiavassa}, {Cocozza}, {Costigan}, {Cowell}, {Crifo}, {Crosta},
  {Crowley}, {Cuypers}, {Dafonte}, {Damerdji}, {Dapergolas}, {David}, {David},
  {de Laverny}, \& {De Luise}}]{Gaia2018}
{Gaia Collaboration}, {Brown}, A.~G.~A., {Vallenari}, A., {et~al.} 2018, \aap,
  616, A1, \dodoi{10.1051/0004-6361/201833051}

\bibitem[{{Garnavich} {et~al.}(2023){Garnavich}, {Wood}, {Milne}, {Jensen},
  {Blakeslee}, {Brown}, {Scolnic}, {Rose}, \& {Brout}}]{garnavich2023}
{Garnavich}, P., {Wood}, C.~M., {Milne}, P., {et~al.} 2023, \apj, 953, 35,
  \dodoi{10.3847/1538-4357/ace04b}

\bibitem[{{Hoyt} {et~al.}(2025){Hoyt}, {Jang}, {Freedman}, {Madore}, {Owens},
  \& {Lee}}]{Hoyt2025}
{Hoyt}, T.~J., {Jang}, I.~S., {Freedman}, W.~L., {et~al.} 2025, arXiv e-prints,
  arXiv:2503.11769, \dodoi{10.48550/arXiv.2503.11769}

\bibitem[{{Jensen} {et~al.}(2024){Jensen}, {Anand}, {Blakeslee}, {Cantiello},
  {Chiboucas}, {Cho}, {Gutcke}, {HyeongHan}, {Jee}, {Kourkchi}, {Lim}, {Mould},
  {Peng}, {Raimondo}, \& {Tully}}]{2024jwst.prop.5989J}
{Jensen}, J., {Anand}, G.~S., {Blakeslee}, J.~P., {et~al.} 2024, {The JWST SBF
  Coma Cluster Survey: Building an Alternative Precision Distance Ladder for
  Cosmology}, JWST Proposal. Cycle 3, ID. \#5989

\bibitem[{{Jensen} {et~al.}(2015){Jensen}, {Blakeslee}, {Gibson}, {Lee},
  {Cantiello}, {Raimondo}, {Boyer}, \& {Cho}}]{jensen2015}
{Jensen}, J.~B., {Blakeslee}, J.~P., {Gibson}, Z., {et~al.} 2015, \apj, 808,
  91, \dodoi{10.1088/0004-637X/808/1/91}

\bibitem[{{Jensen} {et~al.}(2003){Jensen}, {Tonry}, {Barris}, {Thompson},
  {Liu}, {Rieke}, {Ajhar}, \& {Blakeslee}}]{jensen2003}
{Jensen}, J.~B., {Tonry}, J.~L., {Barris}, B.~J., {et~al.} 2003, \apj, 583,
  712, \dodoi{10.1086/345430}

\bibitem[{{Jensen} {et~al.}(2001){Jensen}, {Tonry}, {Thompson}, {Ajhar},
  {Lauer}, {Rieke}, {Postman}, \& {Liu}}]{jensen2001}
{Jensen}, J.~B., {Tonry}, J.~L., {Thompson}, R.~I., {et~al.} 2001, \apj, 550,
  503, \dodoi{10.1086/319819}

\bibitem[{{Jensen} {et~al.}(2021){Jensen}, {Blakeslee}, {Ma}, {Milne}, {Brown},
  {Cantiello}, {Garnavich}, {Greene}, {Lucey}, {Phan}, {Tully}, \&
  {Wood}}]{jensen2021}
{Jensen}, J.~B., {Blakeslee}, J.~P., {Ma}, C.-P., {et~al.} 2021, \apjs, 255,
  21, \dodoi{10.3847/1538-4365/ac01e7}

\bibitem[{Madore \& Freedman(2025)}]{Madore2025}
Madore, B.~F., \& Freedman, W.~L. 2025, The Astrophysical Journal, 983, 161,
  \dodoi{10.3847/1538-4357/adbd3d}

\bibitem[{{Mei} {et~al.}(2007){Mei}, {Blakeslee}, {C{\^o}t{\'e}}, {Tonry},
  {West}, {Ferrarese}, {Jord{\'a}n}, {Peng}, {Anthony}, \& {Merritt}}]{mei2007}
{Mei}, S., {Blakeslee}, J.~P., {C{\^o}t{\'e}}, P., {et~al.} 2007, \apj, 655,
  144, \dodoi{10.1086/509598}

\bibitem[{{Planck Collaboration} {et~al.}(2020){Planck Collaboration},
  {Aghanim}, {Akrami}, {Ashdown}, {Aumont}, {Baccigalupi}, {Ballardini},
  {Banday}, {Barreiro}, {Bartolo}, {Basak}, {Battye}, {Benabed}, {Bernard},
  {Bersanelli}, {Bielewicz}, {Bock}, {Bond}, {Borrill}, {Bouchet}, {Boulanger},
  {Bucher}, {Burigana}, {Butler}, {Calabrese}, {Cardoso}, {Carron},
  {Challinor}, {Chiang}, {Chluba}, {Colombo}, {Combet}, {Contreras}, {Crill},
  {Cuttaia}, {de Bernardis}, {de Zotti}, {Delabrouille}, {Delouis}, {Di
  Valentino}, {Diego}, {Dor{\'e}}, {Douspis}, {Ducout}, {Dupac}, {Dusini},
  {Efstathiou}, {Elsner}, {En{\ss}lin}, {Eriksen}, {Fantaye}, {Farhang},
  {Fergusson}, {Fernandez-Cobos}, {Finelli}, {Forastieri}, {Frailis},
  {Fraisse}, {Franceschi}, {Frolov}, {Galeotta}, {Galli}, {Ganga},
  {G{\'e}nova-Santos}, {Gerbino}, {Ghosh}, {Gonz{\'a}lez-Nuevo}, {G{\'o}rski},
  {Gratton}, {Gruppuso}, {Gudmundsson}, {Hamann}, {Handley}, {Hansen},
  {Herranz}, {Hildebrandt}, {Hivon}, {Huang}, {Jaffe}, {Jones}, {Karakci},
  {Keih{\"a}nen}, {Keskitalo}, {Kiiveri}, {Kim}, {Kisner}, {Knox},
  {Krachmalnicoff}, {Kunz}, {Kurki-Suonio}, {Lagache}, {Lamarre}, {Lasenby},
  {Lattanzi}, {Lawrence}, {Le Jeune}, {Lemos}, {Lesgourgues}, {Levrier},
  {Lewis}, {Liguori}, {Lilje}, {Lilley}, {Lindholm}, {L{\'o}pez-Caniego},
  {Lubin}, {Ma}, {Mac{\'\i}as-P{\'e}rez}, {Maggio}, {Maino}, {Mandolesi},
  {Mangilli}, {Marcos-Caballero}, {Maris}, {Martin}, {Martinelli},
  {Mart{\'\i}nez-Gonz{\'a}lez}, {Matarrese}, {Mauri}, {McEwen}, {Meinhold},
  {Melchiorri}, {Mennella}, {Migliaccio}, {Millea}, {Mitra},
  {Miville-Desch{\^e}nes}, {Molinari}, {Montier}, {Morgante}, {Moss}, {Natoli},
  {N{\o}rgaard-Nielsen}, {Pagano}, {Paoletti}, {Partridge}, {Patanchon},
  {Peiris}, {Perrotta}, {Pettorino}, {Piacentini}, {Polastri}, {Polenta},
  {Puget}, {Rachen}, {Reinecke}, {Remazeilles}, {Renzi}, {Rocha}, {Rosset},
  {Roudier}, {Rubi{\~n}o-Mart{\'\i}n}, {Ruiz-Granados}, {Salvati}, {Sandri},
  {Savelainen}, {Scott}, {Shellard}, {Sirignano}, {Sirri}, {Spencer},
  {Sunyaev}, {Suur-Uski}, {Tauber}, {Tavagnacco}, {Tenti}, {Toffolatti},
  {Tomasi}, {Trombetti}, {Valenziano}, {Valiviita}, {Van Tent}, {Vibert},
  {Vielva}, {Villa}, {Vittorio}, {Wandelt}, {Wehus}, {White}, {White},
  {Zacchei}, \& {Zonca}}]{planck2020}
{Planck Collaboration}, {Aghanim}, N., {Akrami}, Y., {et~al.} 2020, \aap, 641,
  A6, \dodoi{10.1051/0004-6361/201833910}

\bibitem[{{Raimondo}(2009)}]{Raimondo2009}
{Raimondo}, G. 2009, \apj, 700, 1247, \dodoi{10.1088/0004-637X/700/2/1247}

\bibitem[{{Raimondo} {et~al.}(2005){Raimondo}, {Brocato}, {Cantiello}, \&
  {Capaccioli}}]{Raimondo2005}
{Raimondo}, G., {Brocato}, E., {Cantiello}, M., \& {Capaccioli}, M. 2005, \aj,
  130, 2625, \dodoi{10.1086/497591}

\bibitem[{{Reid} {et~al.}(2019){Reid}, {Pesce}, \& {Riess}}]{Reid2019}
{Reid}, M.~J., {Pesce}, D.~W., \& {Riess}, A.~G. 2019, \apjl, 886, L27,
  \dodoi{10.3847/2041-8213/ab552d}

\bibitem[{{Riess} {et~al.}(2022){Riess}, {Yuan}, {Macri}, {Scolnic}, {Brout},
  {Casertano}, {Jones}, {Murakami}, {Anand}, {Breuval}, {Brink}, {Filippenko},
  {Hoffmann}, {Jha}, {D'arcy Kenworthy}, {Mackenty}, {Stahl}, \&
  {Zheng}}]{riess2022}
{Riess}, A.~G., {Yuan}, W., {Macri}, L.~M., {et~al.} 2022, \apjl, 934, L7,
  \dodoi{10.3847/2041-8213/ac5c5b}

\bibitem[{{Riess} {et~al.}(2023){Riess}, {Anand}, {Yuan}, {Casertano},
  {Dolphin}, {Macri}, {Breuval}, {Scolnic}, {Perrin}, \&
  {Anderson}}]{riess2023}
{Riess}, A.~G., {Anand}, G.~S., {Yuan}, W., {et~al.} 2023, \apjl, 956, L18,
  \dodoi{10.3847/2041-8213/acf769}

\bibitem[{{Riess} {et~al.}(2024{\natexlab{a}}){Riess}, {Anand}, {Yuan},
  {Casertano}, {Dolphin}, {Macri}, {Breuval}, {Scolnic}, {Perrin}, \&
  {Anderson}}]{riess2024a}
---. 2024{\natexlab{a}}, \apjl, 962, L17, \dodoi{10.3847/2041-8213/ad1ddd}

\bibitem[{{Riess} {et~al.}(2024{\natexlab{b}}){Riess}, {Scolnic}, {Anand},
  {Breuval}, {Casertano}, {Macri}, {Li}, {Yuan}, {Huang}, {Jha}, {Murakami},
  {Beaton}, {Brout}, {Wu}, {Addison}, {Bennett}, {Anderson}, {Filippenko}, \&
  {Carr}}]{riess2024b}
{Riess}, A.~G., {Scolnic}, D., {Anand}, G.~S., {et~al.} 2024{\natexlab{b}},
  \apj, 977, 120, \dodoi{10.3847/1538-4357/ad8c21}

\bibitem[{{Said} {et~al.}(2024){Said}, {Howlett}, {Davis}, {Lucey}, {Saulder},
  {Douglass}, {Kim}, {Kremin}, {Ross}, {Aldering}, {Aguilar}, {Ahlen},
  {BenZvi}, {Bianchi}, {Brooks}, {Claybaugh}, {Dawson}, {de la Macorra}, {Dey},
  {Doel}, {Fanning}, {Ferraro}, {Font-Ribera}, {Forero-Romero},
  {Gazta{\~n}aga}, {Gontcho}, {Guy}, {Honscheid}, {Kehoe}, {Kisner}, {Lambert},
  {Landriau}, {Le Guillou}, {Manera}, {Meisner}, {Miquel}, {Moustakas},
  {Mu{\~n}oz-Guti{\'e}rrez}, {Myers}, {Nie}, {Palanque-Delabrouille},
  {Percival}, {Prada}, {Rossi}, {Sanchez}, {Schlegel}, {Schubnell}, {Silber},
  {Sprayberry}, {Tarl{\'e}}, {Vargas Magana}, {Weaver}, {Wechsler}, {Zhou}, \&
  {Zou}}]{Said2024}
{Said}, K., {Howlett}, C., {Davis}, T., {et~al.} 2024, arXiv e-prints,
  arXiv:2408.13842, \dodoi{10.48550/arXiv.2408.13842}

\bibitem[{{Schlafly} \& {Finkbeiner}(2011)}]{SF2011}
{Schlafly}, E.~F., \& {Finkbeiner}, D.~P. 2011, \apj, 737, 103,
  \dodoi{10.1088/0004-637X/737/2/103}

\bibitem[{{Scolnic} {et~al.}(2022){Scolnic}, {Brout}, {Carr}, {Riess}, {Davis},
  {Dwomoh}, {Jones}, {Ali}, {Charvu}, {Chen}, {Peterson}, {Popovic}, {Rose},
  {Wood}, {Brown}, {Chambers}, {Coulter}, {Dettman}, {Dimitriadis},
  {Filippenko}, {Foley}, {Jha}, {Kilpatrick}, {Kirshner}, {Pan}, {Rest},
  {Rojas-Bravo}, {Siebert}, {Stahl}, \& {Zheng}}]{pantheonplus}
{Scolnic}, D., {Brout}, D., {Carr}, A., {et~al.} 2022, \apj, 938, 113,
  \dodoi{10.3847/1538-4357/ac8b7a}

\bibitem[{{Scolnic} {et~al.}(2025){Scolnic}, {Riess}, {Murakami}, {Peterson},
  {Brout}, {Acevedo}, {Carreres}, {Jones}, {Said}, {Howlett}, \&
  {Anand}}]{DESI-Scolnic2025}
{Scolnic}, D., {Riess}, A.~G., {Murakami}, Y.~S., {et~al.} 2025, \apjl, 979,
  L9, \dodoi{10.3847/2041-8213/ada0bd}

\bibitem[{{Tonry} \& {Schneider}(1988)}]{tonry1988}
{Tonry}, J., \& {Schneider}, D.~P. 1988, \aj, 96, 807, \dodoi{10.1086/114847}

\bibitem[{{Tonry} {et~al.}(2001){Tonry}, {Dressler}, {Blakeslee}, {Ajhar},
  {Fletcher}, {Luppino}, {Metzger}, \& {Moore}}]{tonry2001}
{Tonry}, J.~L., {Dressler}, A., {Blakeslee}, J.~P., {et~al.} 2001, \apj, 546,
  681, \dodoi{10.1086/318301}

\bibitem[{{Tully}(2015)}]{Tully2015}
{Tully}, R.~B. 2015, \aj, 149, 171, \dodoi{10.1088/0004-6256/149/5/171}

\bibitem[{{Tully} {et~al.}(2023){Tully}, {Anand}, {Blakeslee}, {Cantiello},
  {Jensen}, \& {Raimondo}}]{2023jwst.prop.3055T}
{Tully}, R.~B., {Anand}, G.~S., {Blakeslee}, J.~P., {et~al.} 2023, {A TRGB
  calibration of Surface Brightness Fluctuations}, JWST Proposal. Cycle 2, ID.
  \#3055

\bibitem[{{Tully} {et~al.}(2025){Tully}, {Anand}, {Blakeslee}, {Cantiello},
  {Jensen}, \& {Raimondo}}]{2025jwst.prop.7034T}
---. 2025, {Distance Scale Linkages between JWST Tip of the Red Giant Branch
  and JWST/HST Surface Brightness Fluctuations (and SNIa in E Hosts)}, JWST
  Proposal. Cycle 4, ID. \#7034

\end{thebibliography}
\end{document}